\documentclass[aps,reprint,prd,a4paper,superscriptaddress,nofootinbib,preprintnumbers]{revtex4-1}
\usepackage{hyperref}
\usepackage{amsmath,amssymb}
\usepackage{graphicx}
\usepackage{pgfplots}
\usepackage{bm}
\usepackage[utf8]{inputenc}
\usepackage{xcolor}
\usepackage{chngpage}
\usepackage{longtable}

\begin{document}
\title{Breaks in interstellar spectra of positrons and electrons derived from time-dependent AMS data}
\date{\today}

\author{Andrea Vittino}
\affiliation{Institute for Theoretical Particle Physics and Cosmology (TTK), RWTH Aachen University, 52056 Aachen, Germany}

\author{Philipp Mertsch}
\affiliation{Institute for Theoretical Particle Physics and Cosmology (TTK), RWTH Aachen University, 52056 Aachen, Germany}

\author{Henning Gast}
\affiliation{I. Physics Institute and JARA-FAME, RWTH Aachen University, 52056 Aachen, Germany}

\author{Stefan Schael}
\affiliation{I. Physics Institute and JARA-FAME, RWTH Aachen University, 52056 Aachen, Germany}

\preprint{TTK-19-14}

\begin{abstract}
Until fairly recently, it was widely accepted that local cosmic ray spectra were largely featureless power laws, containing limited information on their acceleration and transport. This viewpoint is currently being revised in the light of evidence for a variety of spectral breaks in the fluxes of cosmic ray nuclei. Here, we focus on cosmic ray electrons and positrons which at the highest energies must be of local origin due to strong radiative losses. We consider a pure diffusion model for their Galactic transport and determine its free parameters by fitting data in a wide energy range: measurements of the interstellar spectrum by Voyager at MeV energies, radio synchrotron data (sensitive to GeV electrons and positrons) and local observations by AMS up to $\sim 1 \, \text{TeV}$. For the first time, we also model the time-dependent fluxes of cosmic ray electrons and positrons at GeV energies recently presented by AMS, treating solar modulation in a simple extension of the widely used force-field approximation. We are able to reproduce all the available measurements to date. Our model of the interstellar spectrum of cosmic ray electrons and positrons requires the presence of a number of spectral breaks, both in the source spectra and the diffusion coefficients. While we remain agnostic as to the origin of these spectral breaks, their presence will inform future models of the microphysics of cosmic ray acceleration and transport.
\end{abstract}

\maketitle
%------------------------------------------------------------------------------------------------------------------------------------------
%------------------------------------------------------------------------------------------------------------------------------------------
%------------------------------------------------------------------------------------------------------------------------------------------

\section{Introduction}

The last decades have witnessed an impressive effort aimed at understanding the acceleration and the transport of Galactic cosmic rays (CRs). On the observational side, a large number of experiments have presented measurements of local fluxes of various CR species and their combined anisotropy. In addition, measurements of the diffuse gamma-ray flux contain information about the CR fluxes in other regions of the Galaxy. The wealth of accurate and diverse data has challenged our understanding of CR origin and propagation. In particular, the long-held wisdom that CR spectra are featureless power laws from GeV to PeV energies had to be revised in the light of a number of spectral breaks observed, most prominently the ``discrepant hardening'' in nuclei fluxes at rigidities of a few hundred $\text{GV}$~\cite{Panov2009,Yoon:2011aa,2011Sci...332...69A,PhysRevLett.114.171103,PhysRevLett.115.211101}. On the modelling side, this requires modifications of the underlying assumptions, in particular on the shapes of source spectra and the rigidity-dependence of the Galactic diffusion coefficient.

CR electrons and positrons are of central importance in investigating the origin of CRs in that at the highest energies they must be necessarily of local origin. For example, in the usually assumed radiation fields (cf.\ e.g.~\cite{Mertsch:2018bqd}), electrons of $1 \, \text{TeV}$ cool in $\sim 3 \times 10^5 \, \text{yr}$ which limits their distances to $\sim 300 \, \text{pc}$. (Here, we have assumed a diffusion coefficient of $10^{29} \text{cm}^2 \, \text{s}^{-1}$.) Specifically, spectral features at hundreds of GeV or higher energies can be related to individual sources of CR electrons and positrons. In addition, CR electrons and positrons in other regions of the Galaxy contribute to the diffuse emission at radio/microwave and gamma-ray wavelengths by synchrotron emission and Inverse Compton scattering, respectively. Understanding the local fluxes is imperative for any global model of CR electrons and positrons. Finally, CR positrons have received heightened attention as a probe for dark matter annihilation or decay. The fluxes of CR electrons and positrons from astrophysical sources constitute an irreducible background for any search of exotic signatures such that precise predictions are required.

In CR electrons and positrons, two features have created most attention over the last decade: First, the positron fraction, i.e.\ the ratio of the positron flux to the sum of electron and positron fluxes, is rising above $\sim 7 \, \text{GeV}$. The existence of such rise, which was already hinted at in the 1 - 50 GeV energy range by the HEAT observations in 1994 \cite{DuVernois_2001,COUTU1999429}, was proven by the PAMELA orbital observatory~\cite{Adriani:2008zr} and later confirmed by AMS at an unprecedented level of precision~\cite{PhysRevLett.114.171103}. Over the years, this excess of high energy positrons has attracted several interpretations mostly in terms of astrophysical mechanisms such as the emission from pulsar wind nebulae ~\cite{Hooper:2008kg,Profumo:2008ms} or the diffusive shock acceleration of positrons produced in spallation reactions occurring inside the shock region of one or more supernova remnants (SNRs)~\cite{2009PhRvL.103e1104B,2009PhRvD..80l3017A}. Numerous interpretations of this anomaly in terms of dark matter annihilation or decay have also been put forward~\cite{Cirelli:2008pk,Cholis:2008hb}, even if it has been  shown that such interpretations can be in strong tension with the constraints that are derived from other dark matter indirect detection channels~\cite{Boudaud:2014dta}. Secondly, a spectral softening was observed in the sum of electron and positron fluxes at $\sim 1 \, \text{TeV}$ by H.E.S.S. observations~\cite{2008PhRvL.101z1104A,Aharonian:2009ah} and recently confirmed by DAMPE~\cite{Ambrosi:2017wek}. Very recently, AMS reported a spectral cut-off in the positrons around $300 \, \text{GeV}$~\cite{PhysRevLett.122.041102}, notably at significant smaller energies than the break in the all-electron spectrum.

At energies below a few tens of GeV, studies of the interstellar spectra are hampered by solar modulation, that is the energy losses and flux suppression due to the interaction of CRs with the solar wind and its frozen-in magnetic field. (See Ref.~\cite{2013LRSP...10....3P} for a review). This modulation is periodic with a primary period of $11 \, \text{years}$. Until recently, the statistics of the experimental data was such that only CR spectra averaged over significant fractions of the $11 \, \text{year}$ period and therefore only studies of the average properties of solar modulation were possible. These time-averaged data could be reasonably well described by the simple and popular force-field model. Recently, the substantial increase in the number of events collected by the detectors has made time-dependent measurements of CR spectra possible. Time-dependent lepton spectra have been released by PAMELA~\cite{Adriani:2015kxa,PhysRevLett.116.241105} and AMS ~\cite{PhysRevLett.121.051102}. In particular, \cite{Adriani:2015kxa} reports the measurement performed by PAMELA of the electron flux in the [70 MeV - 50 GeV] energy range, binned in seven time bins (of around six months each) that cover the solar minimum from July 2006 to December 2009. PAMELA has also presented the positron-to-electron ratio in the [500 MeV - 5 GeV] energy range for 35 time intervals (of around three months each) between July 2006 and December 2015~\cite{PhysRevLett.116.241105}. AMS has presented the electron flux, the positron flux and the positron-to-electron ratio measured in the [1 GeV - 50 GeV] energy range and in the time period from June 2011 to April 2017, binned in time intervals with a duration of one Bartels rotation each~\cite{PhysRevLett.121.051102}.

The study of solar modulation and modelling of the interstellar spectra  does not only benefit from these new time-dependent measurements, but also from the first direct measurements in the interstellar medium. Specifically, the Voyager I spacecraft, launched in 1977, transited the heliopause in 2012 and entered into interstellar space~\cite{2016ApJ...831...18C}. It should be mentioned, in any case, that the Voyager measurements are at significantly lower energies than those at the Earth's position such that the effect of solar modulation cannot be estimated without an extrapolation or better, modelling, of the spectra.

The aim of this paper is to model the electron and positron local interstellar spectra (LIS) over a wide energy range from tens of MeV to $\sim \text{TeV}$. Our model will contain a number of spectral breaks in the source spectra and in the rigidity dependence of the diffusion coefficient. To this end, we will exploit a variety of complementary experimental datasets. We will emphasise which data set requires the introduction of which spectral break. Our analysis will benefit from the recent time-dependent measurements of the electron and positron fluxes performed by AMS~\cite{PhysRevLett.121.051102}. We will illustrate how the effect of solar modulation on these fluxes can be described to a very good extent within the framework of a simple analytical extension of the force-field approximation.

The paper is organised as follows: In Section~\ref{sec:method} we illustrate the main features that characterize our implementation of the acceleration and transport of CR electrons and positrons. In Section \ref{sec:analysis}, we describe the setup of the different analyses that we perform and we discuss our results. Then, in section~\ref{sec:conclusions} we summarise our findings and provide our conclusions.

%------------------------------------------------------------------------------------------------------------------------------------------
%------------------------------------------------------------------------------------------------------------------------------------------
%------------------------------------------------------------------------------------------------------------------------------------------ 

\section{Method} 
\label{sec:method}

\subsection{CR sources}
\label{sec:injection}

CR electrons and positrons can be of either primary or secondary origin.  Primary CRs are those particles that undergo acceleration in astrophysical sources. Primary electrons are expected to be accelerated by SNRs through diffusive shock acceleration. The number of CRs of a given species injected by SNRs into the ISM per unit time, volume and energy is described by a source term that can be expressed as:
\begin{equation}
\mathcal{Q}_{\mathrm{SNR}} = \mathcal{Q}_0 f(r,z) g(\mathcal{R})
\label{eq:source_SNR}
\end{equation}
where we have made the standard assumption that the rigidity and
spatial dependence can be factorized. The rigidity-dependence is
defined by the function $g(\mathcal{R})$ which we assume to be a
power-law, possibly with a number of breaks, as will be illustrated in
greater detail in Section~\ref{sec:analysis} for the different steps
of our analysis. The function $f(r,z)$, which describes the spatial
dependence of the SNR source term, is assumed to be the one proposed
by \cite{2001RvMP...73.1031F}\footnote{We have verified
  that the use of the source profile proposed in \cite{Lorimer:2006qs}
  gives identical results to the ones presented in Section
  \ref{sec:analysis}. On the other hand, the profile proposed in
  \cite{Case:1998qg}, which is significantly different than the ones
  in \cite{2001RvMP...73.1031F} and \cite{Lorimer:2006qs}, results in
  a slightly different $e^\pm$ spectrum in the MeV range and therefore would require different values for the parameters that will be introduced
  to fit Voyager data in Section \ref{sec:1breaks}.}. Lastly, the
normalization factor $\mathcal{Q}_0$ takes into account the rate of supernova explosions and the luminosity that SNRs inject into the ISM in the form of CRs. 

As mentioned above, for the rise in the positron fraction several interpretations have been put forward. In this paper we take a model-independent viewpoint and assume this extra component of high-energy electrons and positrons to have a spatial dependence that traces the one of SNRs and a rigidity dependence that can be expressed as a power-law with an exponential cut-off:
\begin{equation}
\mathcal{Q}_{\mathrm{extra}} = N_x \left(\frac{\mathcal{R}}{\mathcal{R}_0}\right)^{-\Gamma_x} \mathrm{exp}\left( - \frac{\mathcal{R}}{\mathcal{R}_{\mathrm{cut}}}\right) f(r,z).
\label{eq:source_extra}
\end{equation}
Such a spectrum is compatible with models that describe the
acceleration of electrons and positrons in the magnetosphere of
pulsars (see the discussion in \cite{Profumo:2008ms}) as well as with
models that describe the acceleration of secondary positrons in SNRs
(as detailed in \cite{2009PhRvD..80l3017A}). In all our
investigations, we adopt $\mathcal{R}_{\mathrm{cut}}$ = 600 GV.
We assume $\mathcal{Q}_{\mathrm{extra}}$ to be a charge-symmetric
source term, in the sense that the electron and positron
spectra injected into the ISM by the extra source are 
identical. This assumption is not perfectly consistent with our
model-independent take on the extra term. Indeed, while pulsars are
expected to be charge-symmetric sources of CR leptons, other sources
invoked as interpretations to the rising positron fraction may be
not. As an example, if one assumes this extra source to be SNRs
accelerating secondaries produced in spallation reactions, such a
mechanism will produce slightly more positrons than electrons (as a consequence
of charge-conservation in proton-proton collisions). If we were to fix the normalisation of the extra source by fitting to the positron flux, the small charge-asymmetry would have a negligible impact on the electron flux as typically
high-energy electrons are dominated by the SNR component described by
$\mathcal{Q}_{\mathrm{SNR}}$. In any case, one has to consider that within the present experimental precision a charge symmetric source term can neither be confirmed nor excluded.

Secondary electrons and positrons are produced by the interaction of
primary CR (mostly p and He) scattering off the hydrogen and helium
nuclei of the ISM. We describe this process with the source term
\begin{equation}
\begin{aligned}
&\mathcal{Q}_{e^{\pm}}(\mathcal{R}, \vec{x}) = \\ &4 \pi \sum_{i=\mathrm{p,He}} \sum_{j=\mathrm{H,He}} n_j \int dE_i \Phi_i(\mathcal{R},\vec{x})\frac{d\sigma}{d\mathcal{R}}(i+j \rightarrow e^{\pm} + X),
\label{eq:source_sec}
\end{aligned}
\end{equation}
where $\frac{d\sigma}{d\mathcal{R}}(i+j \rightarrow e^{\pm} + X)$
represents the differential inclusive cross section for the production
of electrons and positrons in $ij$ reactions: for the $pp$ case we
take the parametrization proposed in \cite{Kamae:2004xx,Kamae:2006bf},
while for the processes involving He, as a projectile or as a target,
we adopt the rescaling of the $pp$ cross section obtained by following
the prescriptions given in \cite{Norbury:2006hp}. The quantity $n_j$
represents the density of the target species $j$, which is taken from
the model discussed in \cite{Strong:2004de}, while
$\Phi_i(\mathcal{R},\vec{x})$ is the flux of the primary CR species
$i$.

%------------------------------------------------------------------------------------------------------------------------------------------
%------------------------------------------------------------------------------------------------------------------------------------------
\subsection{Galactic propagation setup}
\label{sec:propagation}

CRs propagate across the diffusive halo of the Galaxy, which we assume
here to be a cylinder of half-height $H=4\,\mathrm{kpc}$ and radius
$R=20\,\mathrm{kpc}$. The propagation is characterized by the
interplay between several processes and it is typically described in
terms of a transport equation which models the time evolution of the
CR density per unit momentum $\Psi(\vec{r},p,t)$, which is related to
the CR flux $\Phi$ by the relation $\Phi = v / (4 \pi) \Psi$, with
$v$ being the CR velocity. In full generality, the transport equation
can be written as \cite{1964ocr..book.....G,1990acr..book.....B}:

\begin{equation}
\begin{aligned}
\frac{\partial{\Psi(\vec{r},p,t)}}{\partial t} =  
\mathcal{Q}(\vec{r},p,t) +  \vec{\nabla} \cdot \left( D_{xx} \vec{\nabla}\Psi(\vec{r},p,t) - \vec{V}_c \Psi(\vec{r},p,t)\right) \\ + 
 \frac{\partial}{\partial p} p^2 D_{pp} \frac{\partial}{\partial p} \frac{1}{p^2} \Psi(\vec{r},p,t) - \frac{\partial}{\partial p} \left[ \dot{p} 
 - \frac{p}{3} \left(\vec{\nabla} \cdot \vec{V}_c\right)\Psi(\vec{r,p,t})\right] \\
 - \Psi(\vec{r},p,t) \left( \frac{1}{\tau_f}  - \frac{1}{\tau_r} \right).
 %+ \sum_j \Psi_j (\vec{r},p',t)\left( \frac{1}{\tau^j_f}  - \frac{1}{\tau^j_r} \right), 
\end{aligned}
\label{eq:CRtransport}
 \end{equation}
We adopt the free-escape boundary condition, $\psi(z = \pm H) = 0$. 

 As is customary, we have written the above equation in terms of the CR momentum per nucleon $p$, related to the rigidity $\mathcal{R}$ by the relation $p = (Z/A)\mathcal{R}$, with $A$ and $Z$ being, respectively, the mass and atomic number of the CR species under consideration. The terms on the right-hand-side of eq.~(\ref{eq:CRtransport}) describe, respectively: the
 CR source terms (as described in the previous paragraph), spatial
 diffusion, convection (with the velocity of the convective wind being
 $\vec{V}_c$), diffusive reacceleration, energy losses and, in the
 third line, nuclear fragmentation and radioactive decays. These two
 processes are characterized, respectively, by the timescales $\tau_f$
 and $\tau_r$ (with $\tau_r$ being the lifetime of the CR species
 under consideration). The terms in the third line are to be included
 only when treating CR nuclei, as it is the case discussed in
 Section~\ref{sec:nuclear_fit}.

 In this paper we consider a simplified form of the transport
 equation, which we use to model the transport of all CR species and
 which refers to a scenario where both diffusive reacceleration and
 convection are neglected. Our transport equations, which we solve
 with the publicly available {\tt DRAGON} code \cite{Evoli:2008dv},
 are therefore:

\begin{equation}
\frac{\partial{\Psi}}{\partial t} =  
\begin{cases}
\mathcal{Q} + \vec{\nabla} \cdot D_{xx} \vec{\nabla}\Psi  - \frac{\partial}{\partial p} \dot{p}  - \frac{1}{\tau_f} \Psi - \frac{1}{\tau_r} \Psi & \mathrm{CR}\, \mathrm{nuclei} \, , \\
\mathcal{Q} + \vec{\nabla} \cdot D_{xx} \vec{\nabla}\Psi  - \frac{\partial}{\partial p} \dot{p}  &  \mathrm{CR}\,\mathrm{leptons}. \\
\end{cases}
\label{eq:CRtransport_2}
\end{equation}
In the following, we only consider steady-state solutions, i.e.\ $\partial \Psi / \partial t \equiv 0$.

The energy loss processes that dominate the $\dot{p}$ term are Coulomb
and ionization losses in the case of CR nuclei and synchrotron,
inverse Compton and bremsstrahlung losses in the case of CR
leptons. For a detailed discussion on how these processes can be
modelled, see~\cite{Evoli:2016xgn}. Important
ingredients for the modelling of these energy loss mechanisms are the
gas density, the Galactic magnetic field and the interstellar
radiation field. As already mentioned, for the gas density we follow
the prescriptions of \cite{Strong:2004de}, while the magnetic field
follows the model of \cite{2011ApJ...738..192P} and the interstellar
radiation field is the one described in
\cite{Strong:1998fr,Porter:2005qx}.

The most important mechanism in our modelling of CR transport is
spatial diffusion. CRs diffuse due to their resonant interaction with
turbulent magnetic fields. This diffusion process is treated in terms
of a diffusion coefficient $D_{xx}$, which, in the most general case,
is a tensor whose components are both spatially and
rigidity-dependent. Here we will assume the simplest scenario in which
diffusion is isotropic and homogeneous across the whole diffusive
halo. Hence we consider a scalar diffusion coefficient with no spatial
dependence.

\begin{figure}
\includegraphics[scale=0.30]{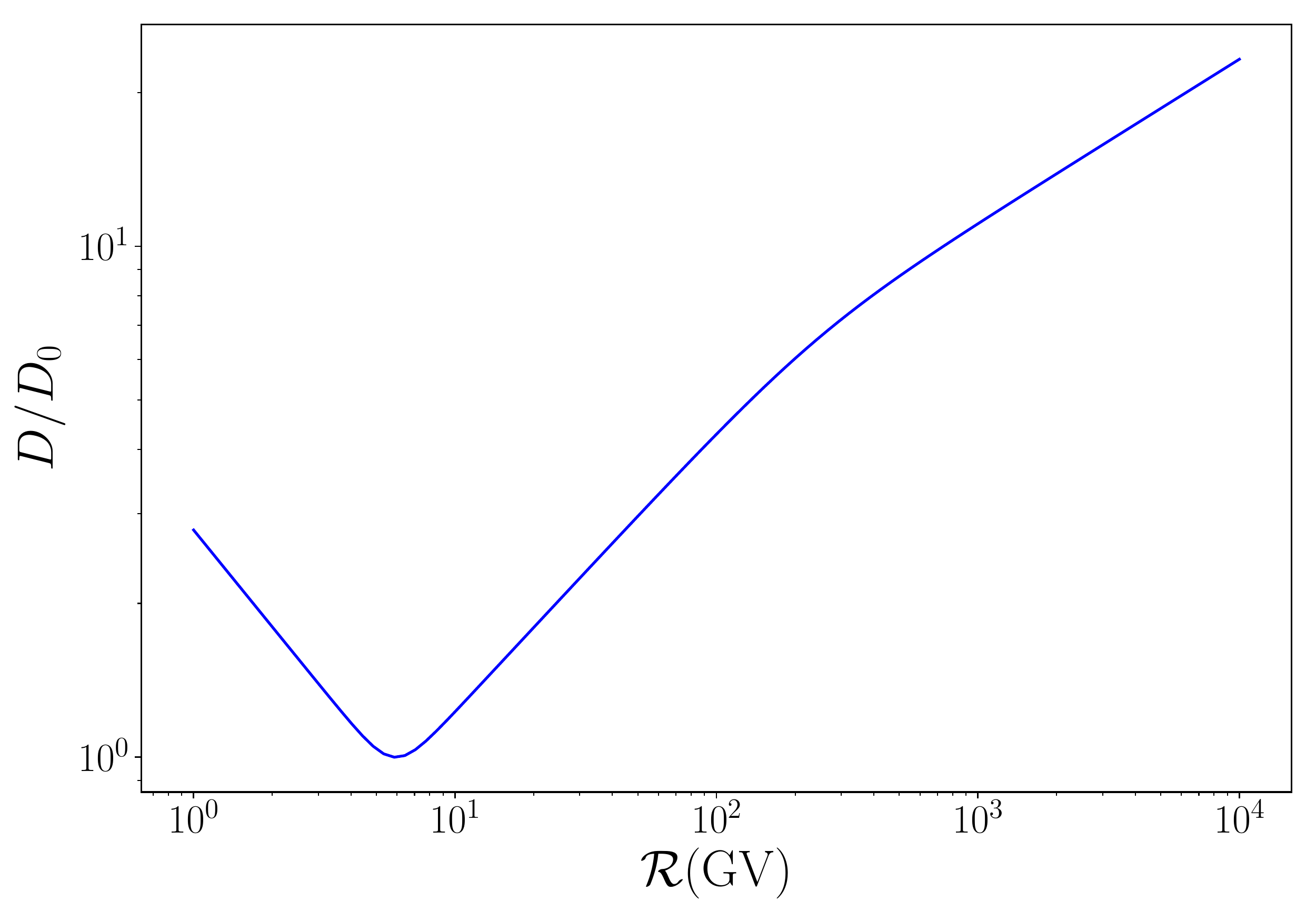}
\caption{Diffusion coefficient as a function of rigidity for the parameters determined below.}
\label{fig:diff_coeff_dbl_broken}
\end{figure}

The rigidity-dependence of the diffusion coefficient is set by the
spectrum of the small-scale turbulence in the interstellar medium. In
particular, a (1D) power law spectrum $k^{-q}$ in wavenumber $k$ leads to a
diffusion coefficient $\propto \mathcal{R}^{\delta}$ where
$\delta = 2 - q$. We assume that the rigidity dependence of the
diffusion coefficient is in the form of an $n$-times broken power law,
\begin{equation}
D_{xx}(\mathcal{R}) = D_0 \beta \left( \frac{\mathcal{R}}{\mathcal{R}_1} \right)^{\delta_1} \prod_{i=1}^n \left( 1 + \left( \frac{\mathcal{R}}{\mathcal{R}_i}\right)^{1/s_i} \right)^{s_i (\delta_{i+1} - \delta_i)} \, ,
\label{eq:diff_coefficient}
\end{equation}

where $D_0$ is a normalization factor, $\beta$ is the velocity of the
particle under consideration, $\delta_i$ are the spectral indices in the rigidity
regimes partitioned by the break rigidities $\mathcal{R}_i$ and the $s_i$ parametrise the smoothness of the  rigidity breaks. For the Galactic diffusion coefficient, we assume two breaks, $n=2$. 
In Fig.~\ref{fig:diff_coeff_dbl_broken}, we show the diffusion coefficient as a function of rigidity with the parameters as determined below.

Under a physical point of view, $\mathcal{R}_1$ (which we assume
to be located at rigidities below 10 GV) is introduced
following~\cite{Ptuskin:2006} to model in an effective way the damping
of turbulence due to an (almost) isotropic distribution of cosmic
rays. Moreover, as it will be detailed in the following, the presence
of such a break is required to reproduce the behaviour of the diffuse
radio emission. On the other hand, the second break
$\mathcal{R}_2 $, which we expect to be at around 200 GV, is
introduced as it provides a satisfactory fit to the most recent data
from AMS \cite{PhysRevLett.120.021101,PhysRevLett.119.251101}, which
have clearly shown that, at this rigidity, the fluxes of light
secondary CRs (Li, Be and B) exhibit identical hardening, stronger
than the one that characterizes the flux of primary CRs (He, C, O). A
possible physical motivation for the existence of a break in the
diffusion coefficient at around 200 GV could be a change in the origin
of the turbulence that is responsible for CR diffusion. As an example,
in \cite{2012PhRvL.109f1101B} it has been suggested that such a break
could be associated to the transition between diffusion in an external
turbulence (as the one injected from SNRs) and diffusion onto CR
self-generated waves (through the mechanism of streaming instability).

%------------------------------------------------------------------------------------------------------------------------------------------
%------------------------------------------------------------------------------------------------------------------------------------------

\subsection{Solar modulation}
\label{sec:solmod}

Before reaching Earth, CRs have to cross the heliosphere. This region
hosts a turbulent magnetic field, together with a hot and ionised
outflow called solar wind. The interaction of CRs with these agents
impacts the CR distribution function, in a process known as solar
modulation. Similarly to the case of Galactic propagation discussed
above, also solar modulation can be modelled by means of a transport
equation. Assuming a steady-state scenario and no injection of CRs in the heliosphere, this equation can be written as \cite{1965P&SS...13....9P}:

\begin{equation}
\vec{V} \cdot \nabla f - \nabla (K \nabla f) - \frac{\mathcal{R}}{3} (\nabla \cdot \vec{V}) \frac{\partial f}{ \partial \mathcal{R}} = 0.
\label{eq:solarmod}
\end{equation}

where $f$ is the CR phase space density (related to the number density
by the relation $d\Psi = f(r,p)d^3r d^3p$), while $K$ is the spatial
diffusion coefficient in the heliosphere, $\vec{V}$ is the velocity of
the solar wind.
Here below we discuss two ways of solving
Eq.~(\ref{eq:solarmod}).

\subsubsection{Standard force-field approximation}
\label{sec:solarmod_standard}

A way of solving eq.~(\ref{eq:solarmod}) is within the framework of
the so-called force-field approximation introduced in
\cite{1968ApJ...154.1011G}. Such a scenario is characterised by a
series of simplifying assumptions. In particular, these assumptions
consist in considering spherically symmetric boundary conditions at the heliospheric radius $R$, 
a radially-directed and
constant solar wind velocity ($\vec{V} = V \hat{r}$) and a uniform and
isotropic spatial diffusion (such that $K$ is a scalar). Moreover, one
has to assume that the CR streaming (or radial current density), under
the influence of diffusion and convection, is zero:

\begin{equation}
C V f- \kappa \frac{\partial f}{\partial r} = 0 \, ,
\end{equation}
where $C = - (1/3) (\partial \ln f)/(\partial \ln p)$ is the Compton-Getting factor.

Under these assumptions, eq.~(\ref{eq:solarmod}) simplifies to
\begin{equation}
\frac{\mathcal{R}V}{3K} \frac{\partial f}{\partial \mathcal{R}} + \frac{\partial f}{\partial r} = 0 \, ,
\label{eqn:characteristic}
\end{equation}
which can be solved with the method of characteristics. If one assumes
$K =K_0 \beta (\mathcal{R}/ \mathcal{R}_1)$, in the relativistic case (where
$\beta \approx 1$), the CR rigidity at the top of the Earth's
atmosphere (i.e.~after solar modulation) is given by:
\begin{equation}
\mathcal{R}_{\mathrm{TOA}} = \mathcal{R}_{\mathrm{LIS}} - \phi
\label{eqn:FF_RTOA_RLIS}
\end{equation}
where $\mathcal{R}_{\mathrm{LIS}}$ is the CR local interstellar
momentum (i.e.~the momentum before solar modulation) and
\begin{equation}
\phi = \frac{V R}{3K_0} \mathcal{R}_1 \,
\label{eqn:FF_phi}
\end{equation}
is the force-field
potential\footnote{It is important to point out that $\phi$ has the
  dimensions of a rigidity only in this specific case where
  $\beta = 1$ and $K \propto \mathcal{R}$.}. Once that the relation
between $\mathcal{R}_{\mathrm{TOA}}$ and $\mathcal{R}_{\mathrm{LIS}}$
is known, one can exploit the fact that the CR distribution function
is conserved (as a consequence of Liouville's theorem) and write the
CR top-of-atmosphere intensity as
\begin{equation}
J_{\mathrm{TOA}} = f \mathcal{R}_{\mathrm{TOA}}^2 = \left( \frac{\mathcal{R}_{\mathrm{TOA}}}{\mathcal{R}_{\mathrm{LIS}}}\right)^2 J_{\mathrm{LIS}}.
\end{equation} 

\subsubsection{Extended force-field approximation}
\label{sec:solarmod_extended}

To investigate the recent AMS time-dependent electron and positron
data, we construct an extension of the force-field approximation,
which is based on assuming a more general rigidity dependence of the
CR diffusion coefficient in the heliosphere. In particular, we assume
a broken power-law behaviour:
\begin{equation}
K(\mathcal{R}) = K_0 \beta \left( \frac{\mathcal{R}}{\mathcal{R}_1} \right)^{\gamma_1} \left( 1 + \left( \frac{\mathcal{R}}{\mathcal{R}_1}\right)^{1/s} \right)^{s (\gamma_2 - \gamma_1)} \, ,
\label{eq:k_extFF}
\end{equation} 

The physical motivation behind this assumption is that the rigidity-dependence of the diffusion coefficient reflects the wavenumber-dependence of the power spectrum of the turbulent component of the magnetic field. For resonant interactions between the CRs and the turbulent magnetic field, there is a one-to-one relation between the particle's rigidity and the turbulence's wavenumber, with the rigidity being inversely proportional to the resonant wavenumber. The range of wavenumbers far above $2 \pi / l_0$, where $l_0$ is the outer scale of turbulence, is referred to as the inertial range and is commonly modelled with a power spectrum $P(k) \propto k^{-q}$, e.g. with $q = 5/3$ for a Kolmogorov phenomenology. For smaller wavenumbers, the turbulent power is usually significantly suppressed. Quasi-linear theory then predicts resonant interactions for particles with rigidities small enough such that the resonant wavenumber is above $2 \pi / l_0$ and a diffusion coefficient $K \propto \mathcal{R}^{2-q}$. For rigidities large enough such that the resonant wavenumber is below $2 \pi / l_0$, interactions are non-resonant and transport is in the small-angle scattering limit with $K \propto \mathcal{R}^{2}$. (See for example the recent discussion presented in \cite{Gruzinov:2018yxz}.) Note that small-angle scattering is oftentimes considered for ultrahigh energy cosmic rays for which the resonant scale would be beyond any conceivable outer scale of turbulence. However, the only scale in the problem is the outer scale of turbulence such that the problem can be easily scaled to environments with a smaller outer scale, like the heliosphere.

We assume $l_0$ to be equal to the
coherence length of the heliospheric field which has been estimated to
be in a range that goes from 0.0079 AU \cite{2005PhRvL..95w1101M} to
0.04 AU \cite{PhysRevLett.82.3444}. Such values roughly correspond to
rigidities in the interval 3--12 GV. As it will be detailed in
Section~\ref{sec:analysis}, in our analysis we will consider both
$\gamma_1$ and $\gamma_2$ as free parameters. The reason is that, as
discussed for example in \cite{refId0}, while the behaviour of the
spectral indices at scales much smaller and much larger than $l_0$ can
be well understood in terms of the considerations illustrated above,
the same cannot be said about the rigidity dependence of the diffusion
tensor across the coherence length, whose shape might depend on the
turbulence model that is adopted and can even have a functional form
that is more complicated than the broken power-law that we are
imposing here. The TOA rigidity is obtained as a function of LIS rigidity by integrating eq.~(\ref{eqn:characteristic}). Due to the complicated functional form of the diffusion coefficient, the relation between $\mathcal{R}_{\text{TOA}}$ and $\mathcal{R}_{\text{LIS}}$ is not as simple as in the original force-field model, cf.\ eq.~(\ref{eqn:FF_RTOA_RLIS}). However, the overall strength of modulation is still determined by the $\phi$-parameter as before, see eq.~(\ref{eqn:FF_phi}).

AMS time-dependent electron and positron data cover a period of six
years within the 24$^{\mathrm{th}}$ solar cycle, from November 2011
(Bartels rotation number 2427) to April 2017 (Bartels rotation number
2506). In modelling the time dependence of the force-field potential,
we consider the fact that this quantity should have 2 minima at the
two extremes of the time interval that is covered by the dataset that
we consider and a maximum that should correspond to the maximum of the
solar activity in the solar cycle that we are considering (roughly
April 2014). We parametrize $\phi(t)$ as the sum of a constant term, a
Lorentzian function and a hyperbolic tangent:

\begin{equation}
\phi(t) = a + \frac{b - a - \frac{c-a}{2}}{1 + \left( \frac{t-t_0}{\tau}\right)^2} + \frac{c - a}{2} \left( 1 + \mathrm{tanh}\left( \frac{t-t_0}{\tau}\right)\right).
\label{eq:phi_timedep}
\end{equation} 

The time coordinate $t$ corresponds to the Bartels rotation number
(which we shift to have the first data point at t = 0). The parameter
$t_0$ identifies the position of the maximum, $\tau$ parametrises the
width of the maximum, $a = \phi(-\infty)$, $b \approx \phi(t_0)$ and
$c = \phi(\infty)$.

%------------------------------------------------------------------------------------------------------------------------------------------
%------------------------------------------------------------------------------------------------------------------------------------------
%------------------------------------------------------------------------------------------------------------------------------------------
\section{Analysis}
\label{sec:analysis}

\begin{table*}[t]
\centering
\scriptsize
\setlength\tabcolsep{2pt}
\caption{Best-fit parameters for the fit to B/C, proton and helium data. The free normalizations of the proton and helium injection spectra (expressed by the parameter $Q_0$ in eq.~(\ref{eq:source_SNR})) is parametrised in terms of the values of the final fluxes $\Phi_p$ and $\Phi_{\mathrm{He}}$ at the reference rigidity $\mathcal{R}_{\text{ref}}\,=30\,\mathrm{GV}$. }
\begin{tabular}{| c | c | c | c | c | c | c | c | c | c | c | c |c | c | c | 
c | c | c | c  |}
\hline
$D_0$ & $\delta_1$ & $\delta_2$ & $\delta_3$ & $\mathcal{R}_1$ & $\mathcal{R}_2$ & $s_1$ & $s_2$ &$\Phi_p(\mathcal{R}_{\text{ref}})$&  $\theta_{p,1}$ & $\theta_{p,2}$ &$\mathcal{R}^{*}_{p}$ & $\Phi_{\mathrm{He}}(\mathcal{R}_{\text{ref}})$& $\theta_{\mathrm{He},1}$ & $\theta_{\mathrm{He},2}$ &$\mathcal{R}^{*}_{\mathrm{He}}$ & $\phi_{\mathrm{B/C}}$ & $\phi_{\mathrm{He}}$ & $\phi_{p}$ \\ 
$10^{28}\,\mathrm{cm^2s^{-1}}$ &  &  &  & ($\mathrm{GV}$) & ($\mathrm{GV}$) &  &  & $((\mathrm{GeV m^{2}s\,sr)^{-1}})$ & & & ($\mathrm{GV}$) &$(\mathrm{((GeV/n)m^{2}s\,sr)^{-1})}$ &   &  &($\mathrm{GV}$) & ($\mathrm{GV}$) & ($\mathrm{GV}$) & ($\mathrm{GV}$) \\
\hline
3.76 & -0.63 & 0.55 & 0.32 & 5.86 & 240.68 & 0.11 & 0.46  & 1.21 &2.89 & 2.38 & 5.92  & 0.53 & 2.63 & 2.29 & 7.52 & 0.72 & 0.60&  0.65 \\
\hline
\end{tabular}
\label{tab:nuclear_fit}
\end{table*}

\begin{table*}[t]
\caption{$\chi^2$ values associated to the different datasets}

\begin{tabular}{|c|c|c|}
\hline
$\chi^2_{\mathrm{B}/\mathrm{C}}$/d.o.f. & $\chi^2_{\mathrm{He}}$/d.o.f.  & $\chi^2_{\mathrm{p}}$/d.o.f \\
\hline
25.61/58 & 61.44/59 & 45.80/55\\
\hline
\end{tabular}
\label{tab:nuclear_fit_chi2}
\end{table*}

\begin{figure*}[t]
\centering
\includegraphics[width = 0.49 \textwidth]{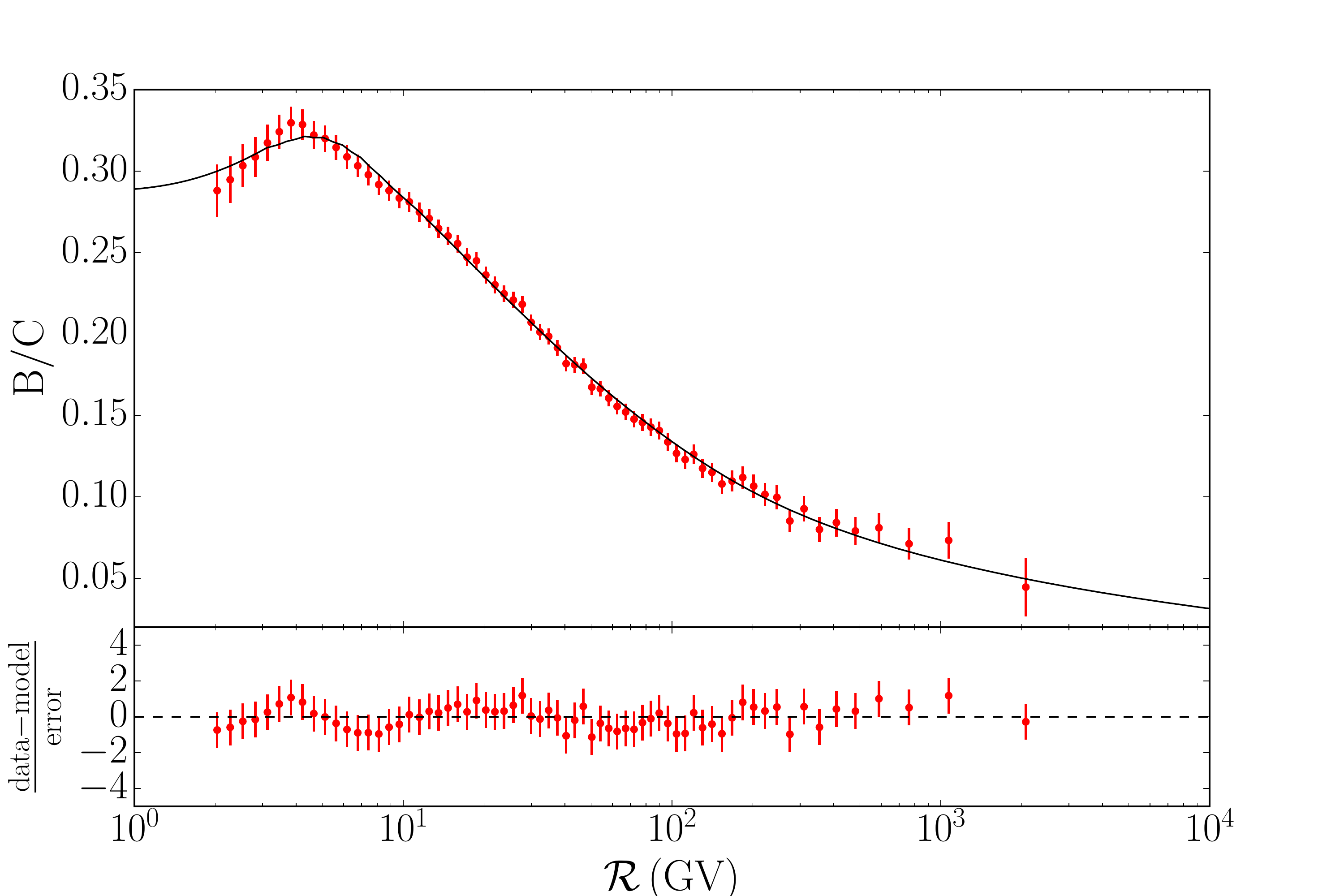}
\includegraphics[width = 0.49 \textwidth]{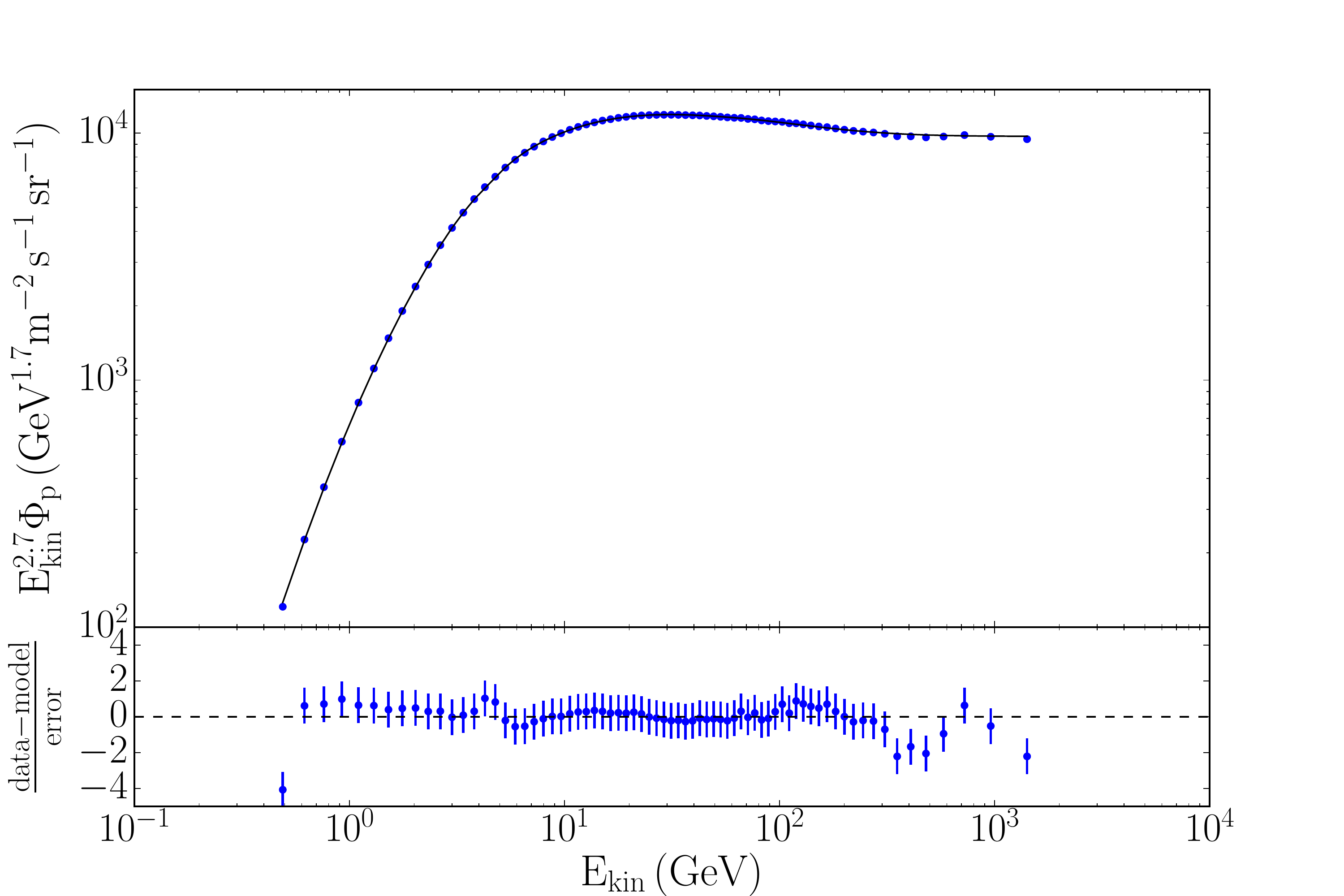}
\caption{The left and right panel show, respectively, the boron-to-carbon ratio and the proton flux for the best-fit configuration of our model, in comparison with AMS data. B/C and proton data are taken, respectively, from \cite{PhysRevLett.120.021101} and \cite{PhysRevLett.114.171103}. For both plots, the lower panels show the pulls. }
\label{fig:nuclear_fit}
\end{figure*}

This section is devoted to the description of the different steps of
our analysis. The analysis consists in fitting models of varying
complexity to different combinations of data sets. Indeed, we start
with a relatively small number of data sets which can be reproduced by
a relatively simple source spectrum. Adding in more data sets requires
more complicated source spectra, and our analysis aims at
understanding which particular data set requires an additional feature
in the source spectrum.

In particular, we start in section \ref{sec:nuclear_fit} by fitting to
the AMS measurement of the boron-to-carbon ratio which fixes the
diffusion coefficient. Next, in section \ref{sec:0breaks} we determine
the source spectrum of electrons and positrons by fitting to the radio
spectrum at high latitudes and the high-energy data on the local
electron and positron spectra as measured by AMS. In section
\ref{sec:1breaks}, we then add the low-energy Voyager I data for the
all-electron spectrum beyond the heliopause, that is the presumed
interstellar flux. Finally, in section \ref{sec:2breaks}, we consider
the recent AMS measurement of the time-dependent electron--positron
data at energies of $1-50 \, \text{GeV}$.

\subsection{Fitting to AMS B/C, proton and helium data}
\label{sec:nuclear_fit}

We fix the parameters $D_0$, $\delta_1$, $\delta_2$, $\delta_3$,
$\mathcal{R}_1$ and $\mathcal{R}_2$ of
Eq.~(\ref{eq:diff_coefficient}) by fitting the AMS data on the
boron-to-carbon ratio (B/C)~\cite{PhysRevLett.120.021101}. In
addition, we fit also the AMS data on CR proton
\cite{PhysRevLett.114.171103} and helium
\cite{PhysRevLett.119.251101}, since the spectra of these CR particle
species are needed to compute the secondary $e^{\pm}$ emission, as
discussed in Section~\ref{sec:injection}.  The systematic
uncertainties of the different data points are treated as detailed in
Appendix~\ref{sec:AMS_uncertainty}.

The proton and helium injected spectra are assumed to be broken power
laws in rigidity. This means that for both species the function
$g(\mathcal{R})$ that appears in Eq.~(\ref{eq:source_SNR}) is assumed
to be of the form:
\begin{equation}
g(\mathcal{R}) =  \left\{ \begin{array}{ll}
\left( \frac{\mathcal{R}}{\mathcal{R}^*} \right)^{-\theta_1} & \text{for } \mathcal{R} \leq \mathcal{R}^{*} \, , \\
\left( \frac{\mathcal{R}}{\mathcal{R}^{*}} \right)^{-\theta_2}  & \text{for } \mathcal{R} > \mathcal{R}^{*} \,  , \\
\end{array} \right.
\end{equation}
where $\mathcal{R}^{*}$ is the position of the rigidity break, where
the spectral index passes from $\theta_1$ to $\theta_2$. For all CR species heavier than helium we adopt the helium
spectral indices.

The proton, helium and B/C data that are considered here in the fit
are \emph{solar modulated} data. As we are mainly concerned with CR
electrons and positrons, we treat solar modulation in terms of the
standard force-field model described in
Section~\ref{sec:solarmod_standard}. In our fit we allow for different
values of the Fisk potential for the different observables under
consideration.

Another remark that has to be made is that when determining the
position of the low-rigidity break in the diffusion coefficient
$\mathcal{R}_1$, we consider only those values of this parameter
for which we are able to find a satisfactory fit to the diffuse
synchrotron emission (such fit will be described in the next Section).

The best-fit parameters found in the fit are reported in Table~\ref{tab:nuclear_fit}, while the $\chi^2$ values associated to the different data sets can be found in Table~\ref{tab:nuclear_fit_chi2} and
the best-fit configurations are shown together with AMS B/C and proton
data in Fig.~{\ref{fig:nuclear_fit}}. The quality of the fit is
remarkably good, as it can be seen, in the case of the proton and B/C
data, from the pulls that are shown in the lower panel of the
plots.

%------------------------------------------------------------------------------------------------------------------------------------------
%------------------------------------------------------------------------------------------------------------------------------------------
\subsection{Fitting to radio and AMS local high-energy electron and positron data}
\label{sec:0breaks}

\begin{table*}[t]
\caption{Best-fit parameters for the three fits to different combinations of data sets considered in our analysis. The parameters $\Phi_{e^-}$ and $\Phi_x$ correspond to the fluxes associated to SNRs and to the extra component at $\mathcal{R} = 30\,\mathrm{GV}$, whose normalizations are a function of the free parameters $N_{e^-}$ and $N_x$ respectively. The uncertainty on each parameter is determined through the procedure described in Appendix~\ref{sec:AMS_uncertainty}.}
\footnotesize
\begin{tabular}{| c | c | c | c |  c | c | c | c | c | c |}
\hline
Model & $\Phi_{e^-}$& $\Phi_{\mathrm{x}}$ & $\Gamma_{\mathrm{x}}$ & $\mathcal{R}_a$ & $\mathcal{R}_b$ & $\alpha_1$ & $\alpha_2$ & $\alpha_3$ & $f_{B}$\\
& $\mathrm{(GeV^{-1} m^{-2} sr^{-1} s^{-1})}$& $\mathrm{(GeV^{-1} m^{-2} sr^{-1} s^{-1})}$ & & $\mathrm{(GV)}$ & $\mathrm{(GV)}$ &  & & & \\
\hline
{\it 0 breaks}& ($4.21 \pm 0.05) \times 10 ^{-3}$& ($2.38 \pm 0.02)\times 10 ^{-4}$ & $1.60 \pm 0.01$ &  - & - & - & $2.58 \pm 0.01$ & - & $3.08^{+0.02}_{-0.05}$ \\
{\it 1 break} & ($4.21 \pm 0.05) \times 10 ^{-3}$& ($2.37 \pm 0.01)\times 10 ^{-4}$ & $1.60^{+0.01}_{-0.03}$ & $0.109^{+0.005}_{-0.003}$ & - & $2.13^{+0.01}_{-0.02}$& $2.57 \pm 0.02$ & - & $3.07^{+0.03}_{-0.08}$ \\
{\it 2 breaks} &
$4.37^{+0.01}_{-0.06}\times 10 ^{-3}$ &$2.39 ^{+0.06}_{-0.02}\times 10 ^{-4}$ & $1.63 \pm 0.01$ & $0.411 ^{+0.001}_{-0.004}$& $83.4^{+1.8}_{-0.1} $&$2.12^{+0.002}_{-0.012}$ &$2.69 \pm 0.01$ &$2.53 \pm 0.01$ & $2.48^{+0.04}_{-0.01}$ \\

\hline
\end{tabular}
\label{tab:bestfit_leptons}
\end{table*}
\begin{table*}[t]
\caption{$\chi^2$ values associated to the fit of the different models to the data sets considered in our analysis.}
\begin{tabular}{| c | c | c | c | c | c | c | c | }
\hline
Model & ${\chi^2_{e^-}}^{\mathrm{HE}}$/d.o.f. & ${\chi^2_{e^+}}^{\mathrm{HE}}$/d.o.f. & $\chi^2_{\mathrm{radio}}$/d.o.f. & $\chi^2_{\mathrm{Voy}}$/d.o.f. & ${\chi^2_{e^-}}^{\mathrm{TD}}$/d.o.f. & ${\chi^2_{e^+}}^{\mathrm{TD}}$ /d.o.f.&
${\chi^2_{e^+/e^-}}^{\mathrm{TD}}$ /d.o.f.\\
\hline
 {\it 0 breaks} & 27.5/20 & 23.0/23 & 3.3/4 & - & - & - & - \\
 {\it 1 break} & 28.0/20 & 23.0/23 & 3.2/4 & 12.1/6 & - & - & - \\
 {\it 2 breaks} & 17.9/18 & 26.3/23 & 6.5/4 &  12.9/6 & 21654/3808 & 6533/3812 & 4510/3798 \\
 \hline
\end{tabular}
\label{tab:chi2_leptons}
\end{table*}

\begin{figure*}[t]
\centering
\includegraphics[width = 0.40 \textwidth]{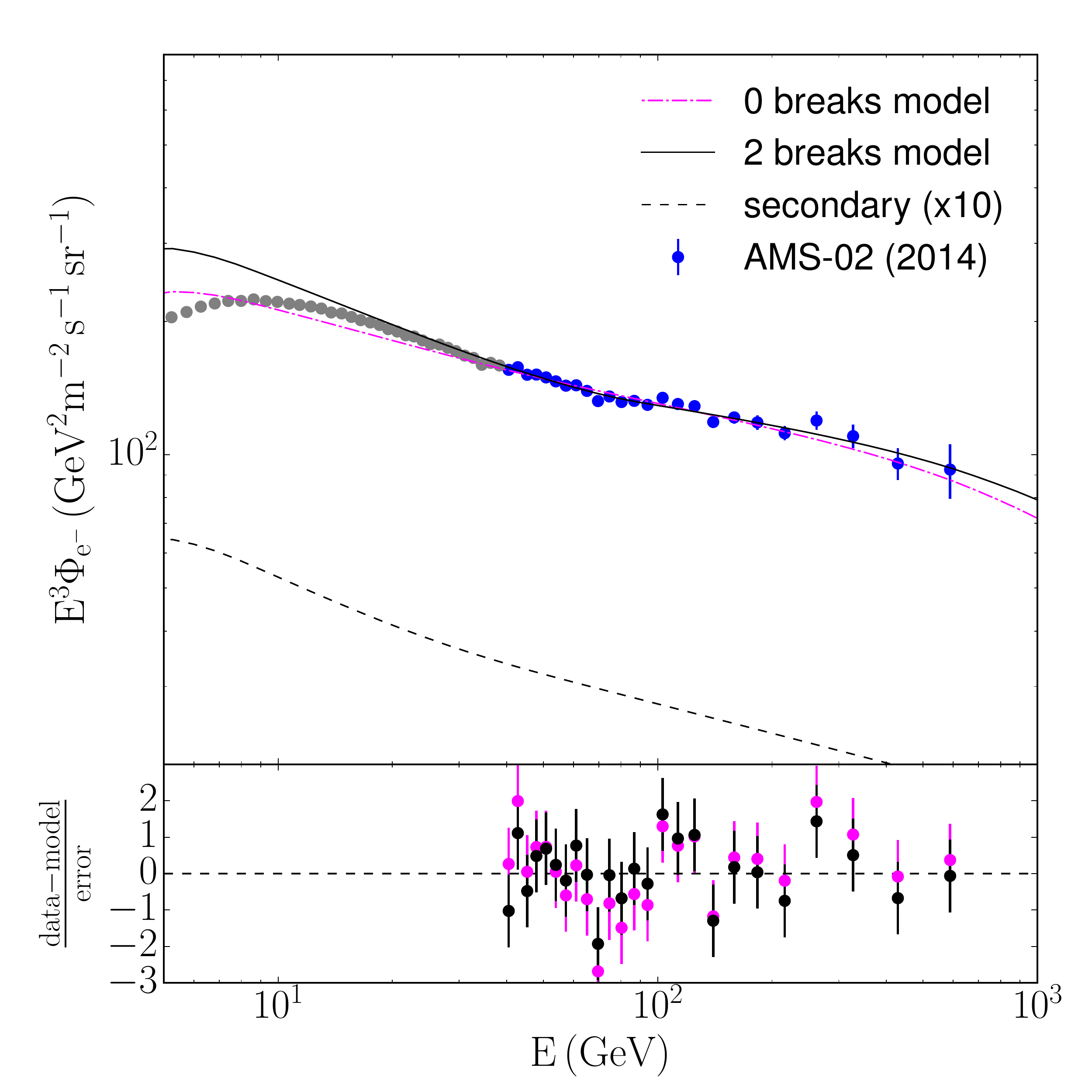}
\includegraphics[width = 0.40 \textwidth]{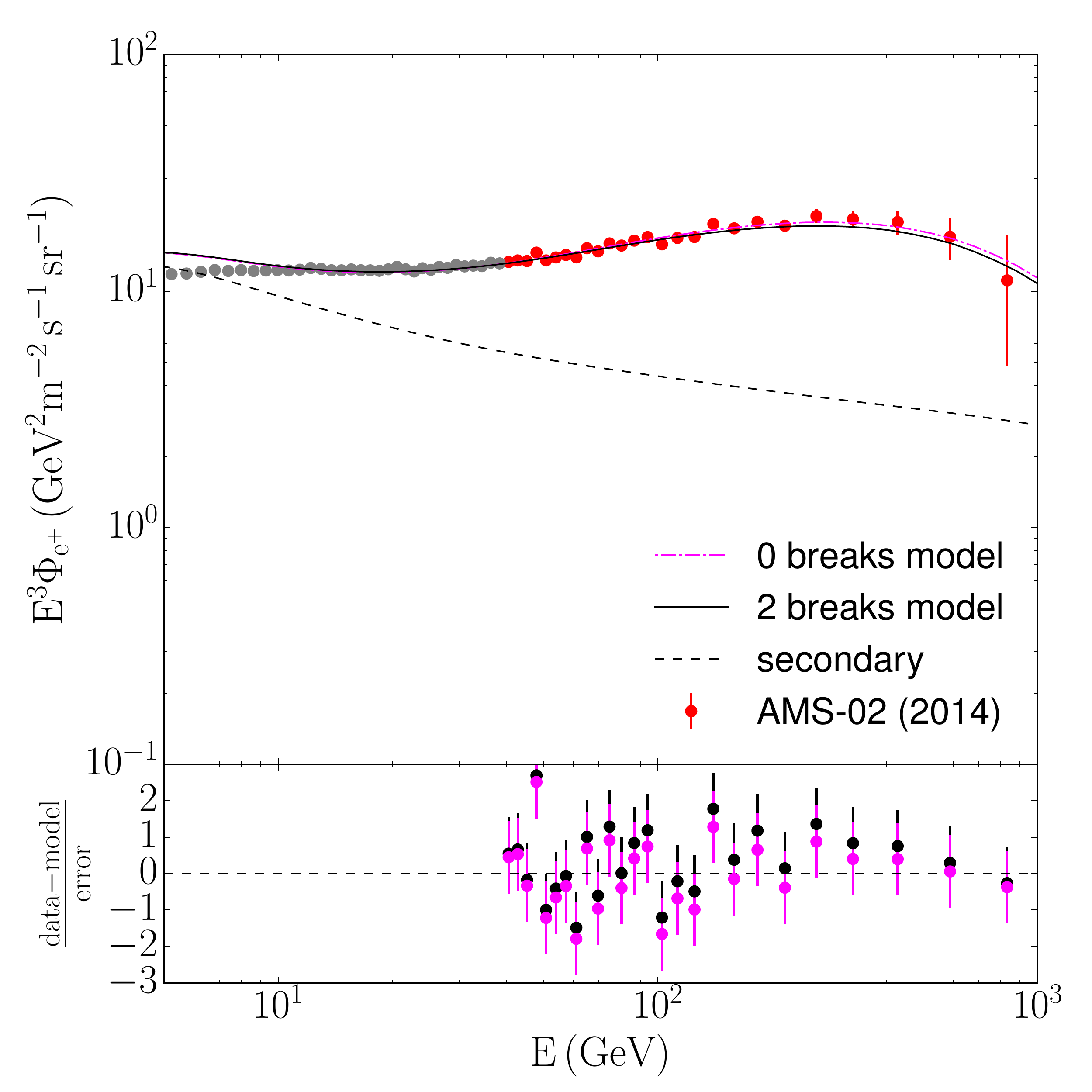}\\
\includegraphics[width = 0.40 \textwidth]{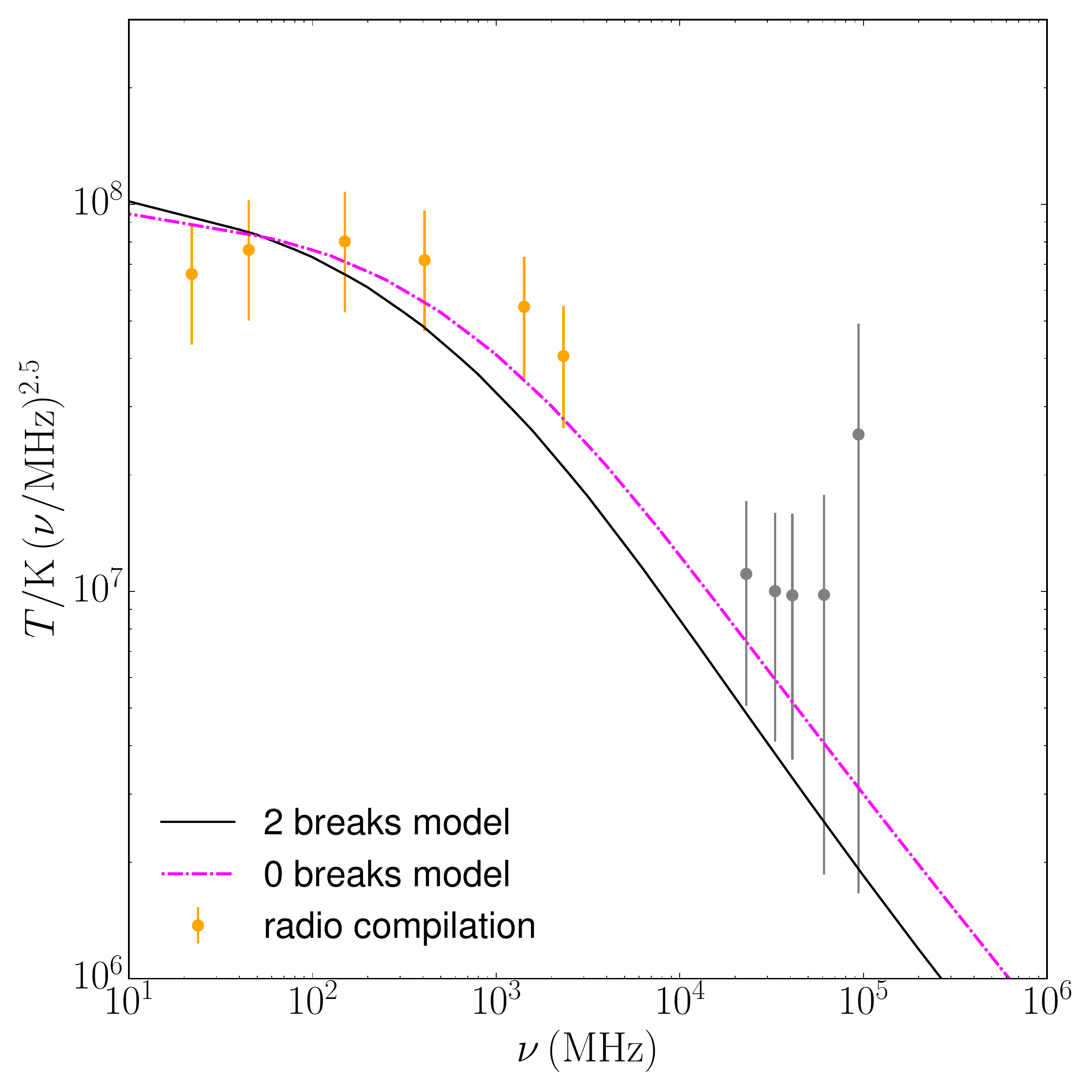}
\includegraphics[width = 0.40 \textwidth]{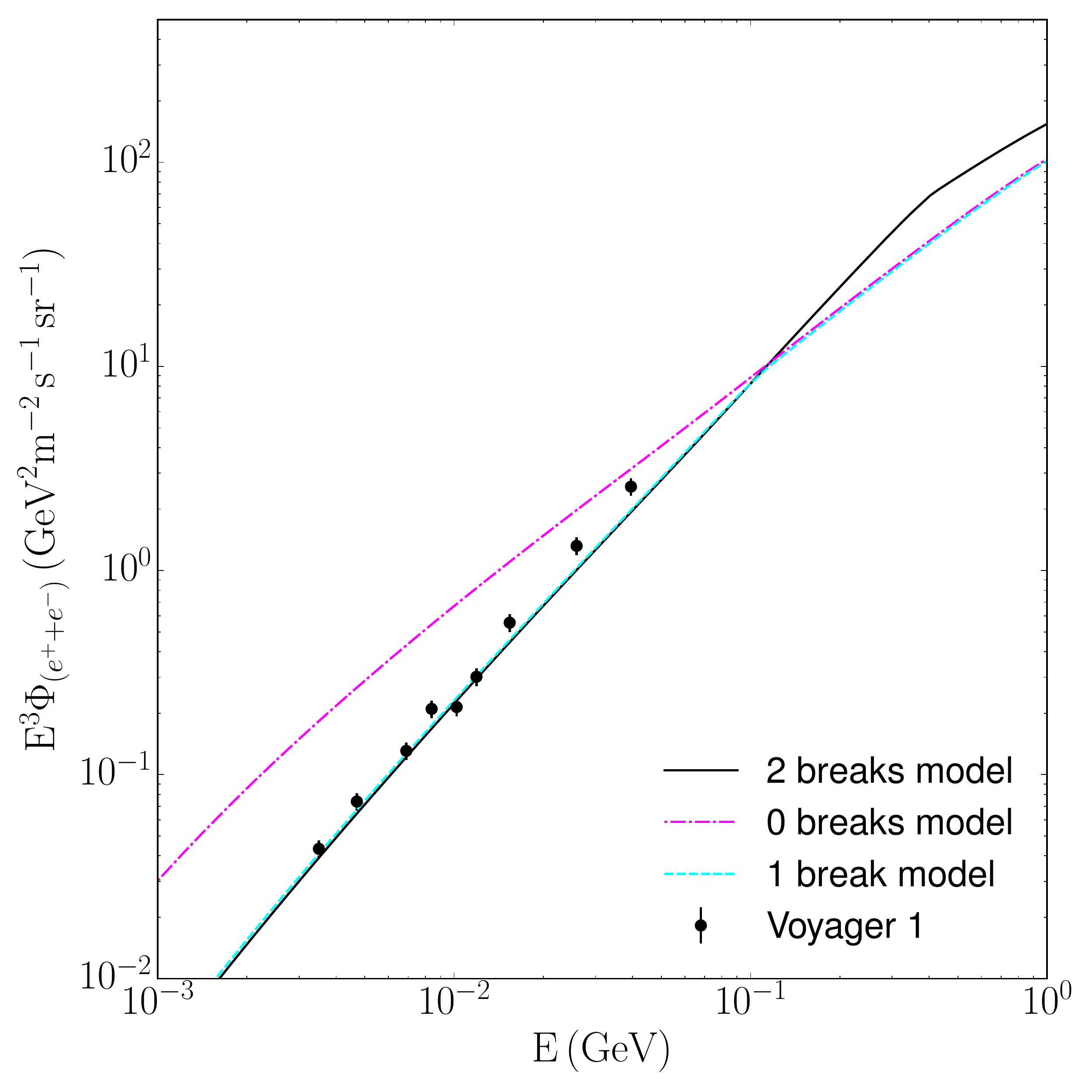}
\caption{In the top row, the electron (left panel) and positron (right
  panel) LIS predicted by the {\it 0 breaks} (magenta dot-dashed lines) and {\it 2 breaks} (black solid lines) models
  described in the text are shown together with AMS data \cite{2014PhRvL.113l1102A,PhysRevLett.122.041102}. We also show the secondary component only (dashed black lines) which is the same in both models. In each
  plot, gray data points are the ones that have been excluded in the
  fitting procedure. The panels below each plot show the pulls (data -
  model/ $\sigma_{\mathrm{exp}}$). The
  bottom left panel illustrates the behaviour of the predicted
  synchrotron emission, for both the {\it 0 breaks} and the {\it 2 breaks}
  models in comparison with data. Data points that are not used in the
  fit are in gray. In the bottom right panel, the all-electron
  flux predicted at low energies by the two models and by the
  {\it 1~break model} is shown together with data from Voyager 1 \cite{2016ApJ...831...18C}. }
\label{fig:results_fit}
\end{figure*}

CR electrons and positrons can be injected in the ISM through a
variety of processes as discussed in Section~\ref{sec:injection}.
Here we assume the source term of primary electrons injected by SNRs
to be a simple power-law, i.e.~the rigidity-dependence of
Eq.~(\ref{eq:source_SNR}) is given by
\begin{equation}
g(\mathcal{R}) = N_{e^-} \left(\frac{\mathcal{R}}{\mathcal{R}_0}\right)^{-\alpha_2}
.
\label{eq:source_SNRs_0breaks}
\end{equation}
As discussed in
Section~\ref{sec:injection}, high-energy positrons and electrons are
assumed to be accelerated by an extra source, with a rigidity spectrum
that depends on two parameters: the normalisation $N_{\mathrm{x}}$ and
the spectral index $\Gamma_{\mathrm{x}}$. Concerning the secondary
component produced by spallation processes, it is modelled as in
Eq.~(\ref{eq:source_sec}), with the proton and helium spectra
determined through the fitting procedure described in the previous
paragraph.

With all this considered, the model that we are investigating here,
which we label {\it 0 breaks model}, is characterised by 5 free
parameters. Four of these
parameters are directly associated to the electrons and positrons source terms: They are the spectral index $\alpha_2$ and the normalization $N_{e^-}$ of the electron
spectrum injected by SNRs (see eq.~(\ref{eq:source_SNRs_0breaks})), the spectral index
$\Gamma_{\mathrm{x}}$ of the $e^\pm$ spectrum injected by the extra component and its normalisation
$N_{\mathrm{x}}$ (see eq.~(\ref{eq:source_extra})). In addition, the synchrotron flux depends on the
strength of the magnetic field $B(\vec{r})$ which varies as a function of position. We adopt a model, with a simple exponential dependence, on the galacto-centric radius $r$ and the distance from the disk, $z$,
\begin{equation}
B(r, z) = B_0 \exp[-(r-r_0)/\rho - |z|/z_0] \, .
\end{equation}
Our results are rather insensitive to the specific values of $r_0$, $\rho$ and $z_0$, and so we adopt $r_0 = 8.5 \, \text{kpc}$, $\rho = 8 \, \text{kpc}$ and $z_0 = 2 \, \text{kpc}$. However, we allow for the normalisation to float in the fit by a factor $f_B$ with respect to the fiducial value of $B_0 = 3 \mu\text{G}$. We expect for $f_B$ to differ from $1$ by a factor of a few at most.

We determine the five free
parameters by fitting to radio and AMS local high-energy electron and
positron data \cite{2014PhRvL.113l1102A,PhysRevLett.122.041102}. In building the data set of radio measurements to be used in the fit,
we follow the approach described in
\cite{Jaffe:2011qw,2011A&A...534A..54S,DiBernardo:2012zu,Orlando:2013ysa,Orlando:2017mvd},
which consists of considering those radio surveys that display a
complete (or nearly complete) sky coverage at several frequencies in
the MHz-GHz interval (more precisely, 22 MHz, 45 MHz, 408 MHz, 1420
MHz, 2326 MHz, 23 GHz, 33 GHz, 41 GHz, 61 GHz and 94 GHz). For each
one of these frequencies, an average flux is estimated by integrating
the sky maps produced by the respective survey over the high-latitude
region, once that the contribution from the Galactic plane and from
radio sources is removed through the application of the WMAP extended
temperature analysis mask \cite{2013ApJS..208...19H}. The uncertainty
associated to the radio flux in each frequency bin is the result of
the variance of the flux in the region of the sky under
consideration. It is important to point out that in our analysis of
the synchrotron data, not all the frequencies enter in the calculation
of the $\chi^2$. In particular, we will not include in our assessment
of the goodness of the fit the frequencies above 10 GHz as at these
frequencies the radio emission is expected to receive important
contributions from free-free and thermal dust emission. Therefore,
one can consider the radio flux determined at these frequencies as an
upper limit to the diffuse emission from CR electrons and positrons.

When fitting to AMS high-energy data, we adopt the prescription
described in Appendix~\ref{sec:AMS_uncertainty} to treat systematic
uncertainties and to determine an uncertainty on the best-fit
parameters. Furthermore, we consider only measurements above a minimum
energy, which we set at $E_{\mathrm{cut}} = 40$ GeV. The reason for
this choice is that we compare AMS data to the unmodulated Local
Interstellar Spectrum (LIS). This means that we have to consider
energies that are sufficiently large that the effect of solar
modulation can be considered negligible with respect to the accuracy
of the data. We have checked that alternative choices of
$E_{\mathrm{cut}}$ do not change our results in a significant way. More details on the impact that solar modulation has on the flux at different energies will be provided in Section~\ref{sec:2breaks} when we will describe the fit to the AMS time-dependent data.   

Within the scenario that we have described in the previous section,
characterised by a double break in the rigidity dependence of the
spatial diffusion coefficient, this simple model is able to fit
remarkably well all the data sets we are considering. Our best-fit
parameters are reported in the first row of
Table~\ref{tab:bestfit_leptons}, while the values of the $\chi^2$
associated to each data set are reported in the first row of
Table~\ref{tab:chi2_leptons}. The best fit configuration is shown in
comparison with data in the upper line and lower left panels of Fig.~\ref{fig:results_fit} (specifically, the {\it 0 breaks model} is represented by the magenta dot-dashed lines).  One important
point to remark is that the fit requires the slope of the primary
electron spectrum to be rather hard and this has a great impact at low
energies, where, as it can be seen in the top left panel of
Fig.~\ref{fig:results_fit}, the electron LIS can even be \emph{below}
the data (in particular in the region below 40 GeV, not included in
the fit). This will pose issues when the low-energy data (and solar
modulation) will be taken into account, as we will discuss in detail
later.

%------------------------------------------------------------------------------------------------------------------------------------------
%------------------------------------------------------------------------------------------------------------------------------------------
\subsection{Fitting to radio, high-energy and Voyager data}
\label{sec:1breaks}

Our intent in this part of the analysis is to constrain the very low
energy range (i.e., below 100 MeV) of the electron LIS. To this end,
we consider the measurements of the total electron flux made by
Voyager 1 \cite{2016ApJ...831...18C} at energies between 2.7 and 74.1
MeV.

As seen in the bottom right panel of Fig.~\ref{fig:results_fit}, the
model, which was found in the previous part of the analysis to provide
a remarkably good fit to high-energy and radio data, produces a total
electron LIS that is in disagreement with Voyager data. More
precisely, the spectrum appears to be too soft and its normalisation
seems too large. We are thus led to adopt a break also in the source
spectrum of primary electrons,

\begin{equation}
g(\mathcal{R}) =  \left\{ \begin{array}{ll}
\left( \frac{\mathcal{R}}{\mathcal{R}^*} \right)^{-\alpha_1} & \text{for } \mathcal{R} \leq \mathcal{R}_{a} \, , \\
\left( \frac{\mathcal{R}_a}{\mathcal{R}^*} \right)^{-\alpha_1} \left( \frac{\mathcal{R}}{\mathcal{R}_{a}} \right)^{-\alpha_2}  & \text{for } \mathcal{R} > \mathcal{R}_{a} \, . \\
\end{array} \right.
\end{equation}
The model that we are considering here, which we label {\it 1 break
  model}, consists of 7 free parameters, which are the 5 that
characterised the {\it 0 breaks model}, plus $\mathcal{R}_a$ and
$\alpha_1$. The best fit parameters and the $\chi^2$ associated to
each data set are reported, respectively, in
Table~\ref{tab:bestfit_leptons} and \ref{tab:chi2_leptons} and the
low-energy total electron LIS is shown, together with Voyager data in
the bottom right panel of Fig.~\ref{fig:results_fit} (specifically, the {\it 1 break model} is represented by the dashed cyan line). The goodness of
fit with respect to this dataset has significantly improved.

%------------------------------------------------------------------------------------------------------------------------------------------
%------------------------------------------------------------------------------------------------------------------------------------------
\subsection{Fitting to radio, high-energy, Voyager and AMS time-dependent data}
\label{sec:2breaks}

In the previous sections, we have investigated the electron and
positron LIS. We now turn our attention to the investigation of the
\emph{solar modulated} fluxes. The idea is that modulated fluxes can
provide additional constraints to our models of the LIS. As
discussed above, for example, the fit to radio data requires a rather
hard electron spectrum which at energies $\mathcal{O}$ (1-10) GeV
might cause the LIS to even be lower than the measured spectrum, thus
leaving very little room, or even no room at all, for solar
modulation. A fit including low-energy local measurements is mandatory
in order to assess this potential issue.

\begin{table*}[t]
\caption{Best-fit parameters for the solar modulation parameters (see Eqs.~(\ref{eq:k_extFF}) and (\ref{eq:phi_timedep})).}
\footnotesize
\begin{tabular}{| c | c | c | c |  c | c | c | c | c |}
\hline
Particle & $\gamma_1$& $\gamma_2$ & $\mathcal{R}_1$ [GV] & $a$ [GV] & $b$ [GV] & $c$ [GV] & $t_0$ [Bartels rot.] & $\tau$ [Bartels rot.]  \\
\hline
electron & $1.32^{+0.05}_{-0.03}$ & $1.76^{+0.10}_{-0.01}$ &$5.71^{+0.26}_{-0.35}$& $0.12^{+0.02}_{-0.03}$ & $0.585^{+0.016}_{-0.011}$& $0.067^{+0.027}_{-0.035}$ & $2474.54 \pm 0.01$ & $34.01^{+0.74}_{-1.47}$ \\
\hline
positron & $1.25^{+0.01}_{-0.05}$ & $2.29^{+0.29}_{-0.33}$ &$5.13^{+0.52}_{-0.93}$& $0.12^{+0.01}_{-0.04}$ & $0.413^{+0.031}_{-0.011}$& $0.010^{+0.002}_{-0.03}$ & $2467.5 \pm 0.7$ & $26.12^{+3.01}_{-0.83}$ \\
\hline
\end{tabular}
\label{tab:solar}
\end{table*}

\begin{figure*}[t]
\centering
\includegraphics[width = \textwidth]{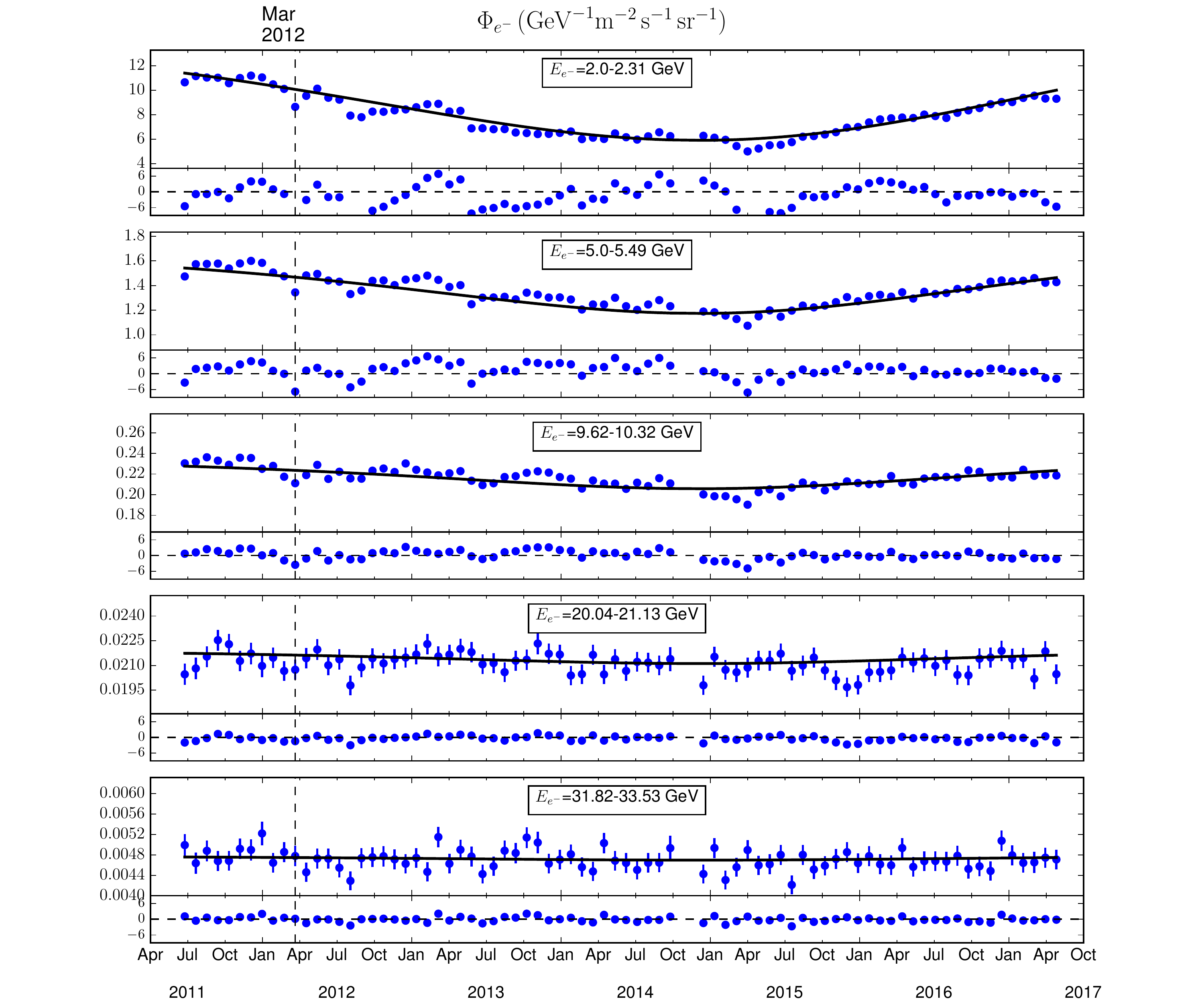}
\caption{Electron time-dependent fluxes, as predicted by our model
  (solid lines) and as measured by AMS \cite{PhysRevLett.121.051102} (points). Each panel refers to
  a specific energy (as reported in the labels).}
\label{fig:ele_solarmod}
\end{figure*}

\begin{figure*}
\centering
\includegraphics[width = \textwidth]{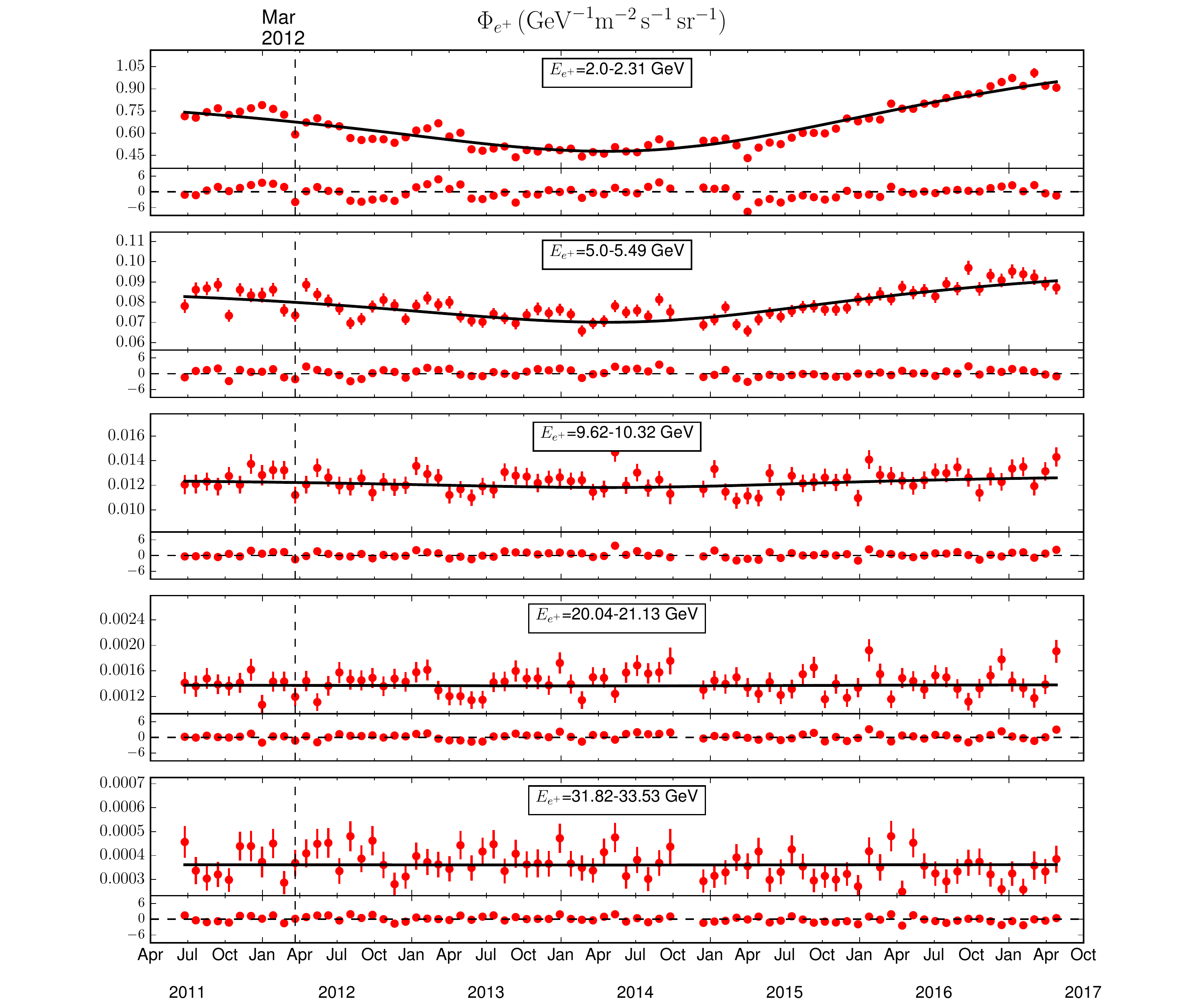}
\caption{Positron time-dependent fluxes, as predicted by our model
  (solid lines) and as measured by AMS \cite{PhysRevLett.121.051102} (points). Each panel refers to
  a specific energy (as reported in the labels).}
\label{fig:pos_solarmod}
\end{figure*}

\begin{figure*}[t]
\centering
\includegraphics[width = \textwidth]{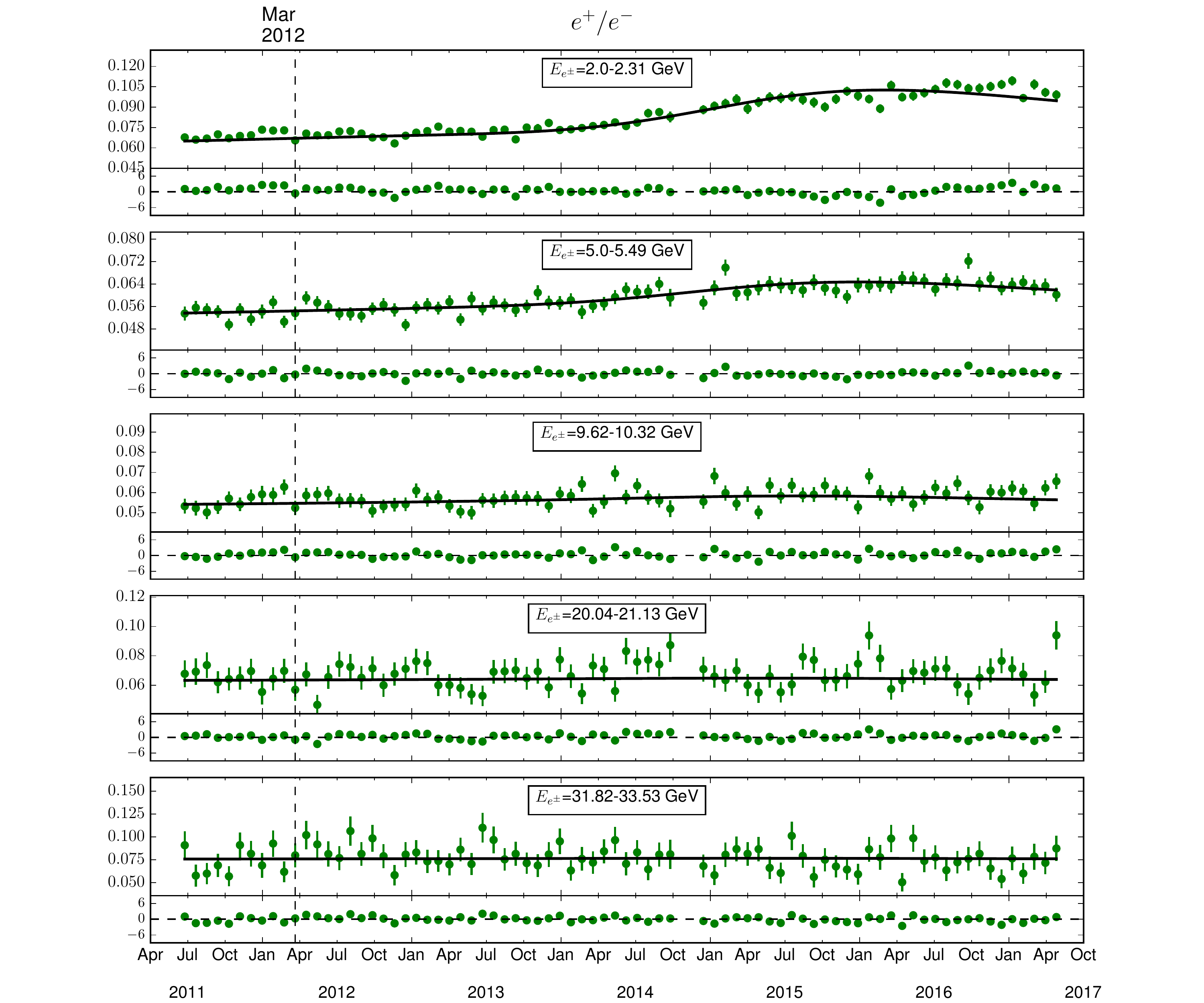}
\caption{Positron-to-electron time-dependent ratio, as predicted by
  our model (solid lines) and as measured by AMS \cite{PhysRevLett.121.051102}(points). Each panel
  refers to a specific energy (as reported in the labels).}
\label{fig:ratio_solarmod}
\end{figure*}

\begin{figure*}[t]
\centering
\includegraphics[width = 0.45\textwidth]{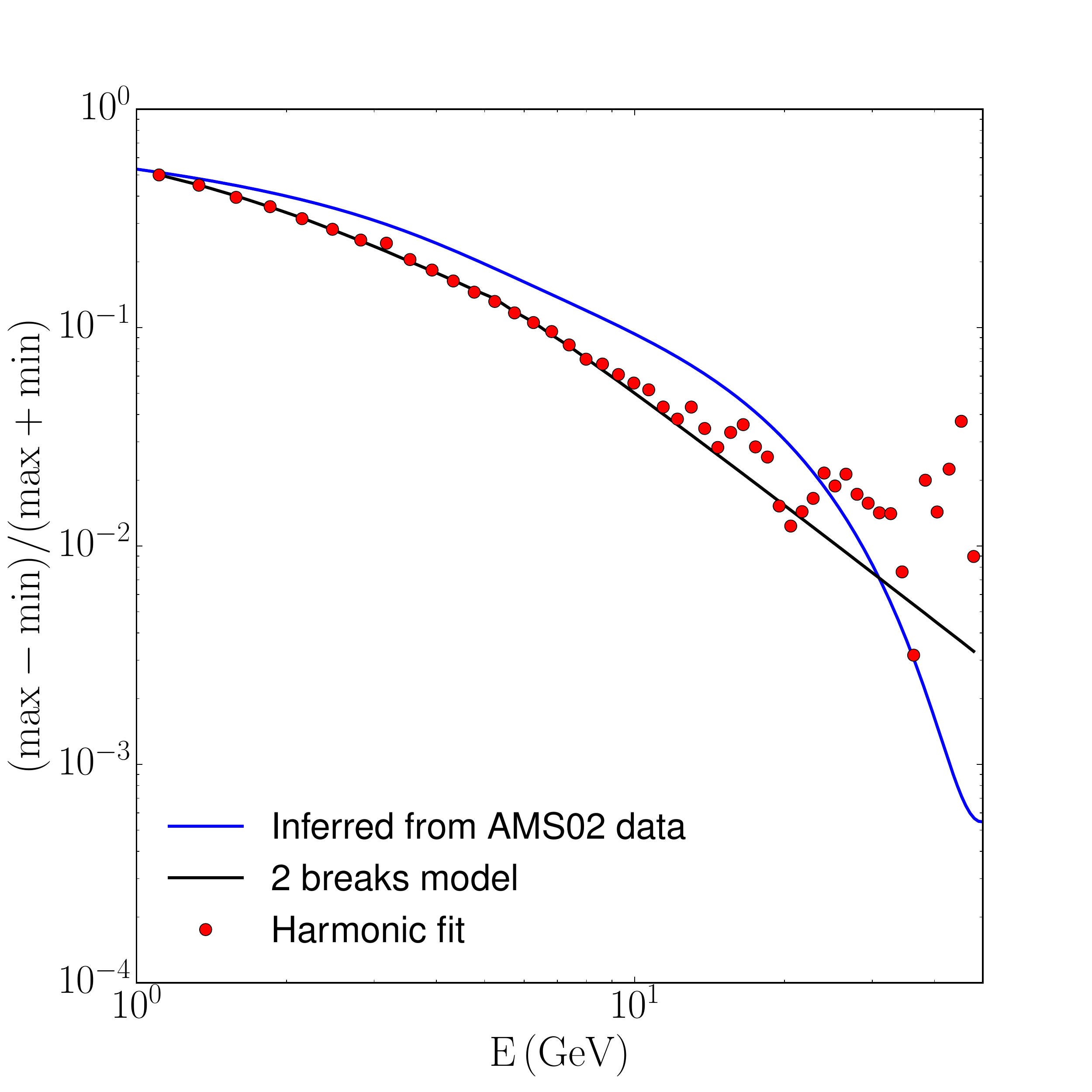}
\includegraphics[width = 0.45\textwidth]{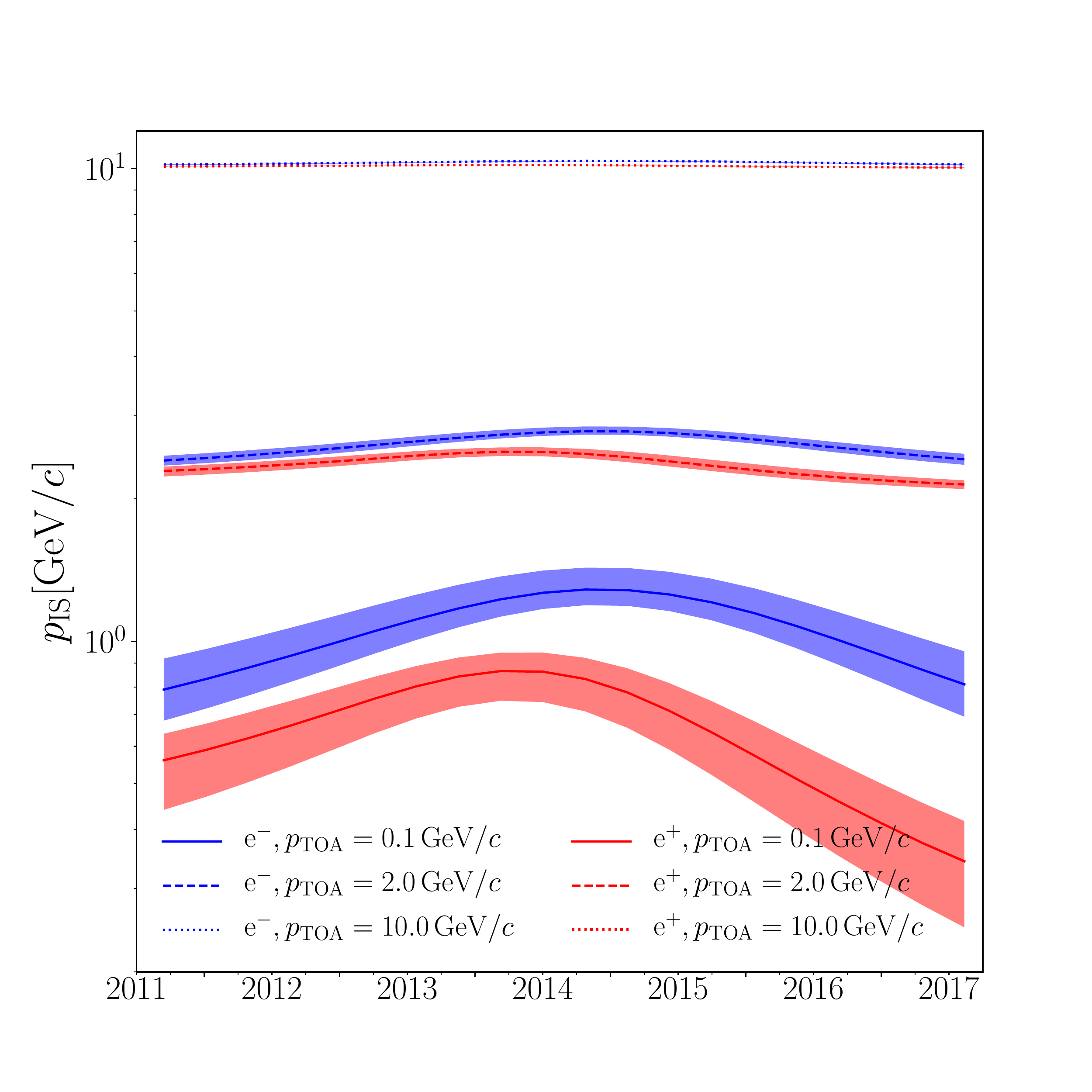}
\caption{\textbf{Left:} Maximum fluctuation of the electron spectrum as a function of
  energy, as inferred from AMS data (blue curve) and as predicted
  within the framework of our solar modulation model (solid black
  line) or by means of the fit with the harmonic function of
  Eq.~(\ref{eq:harmonic}) (red points). \textbf{Right:} Unmodulated momentum $p_{\text{IS}}$ as a function of time for modulated momenta $p_{\text{TOA}} = 0.1$, $2$ and $10 \, \text{GeV}/c$. The shaded bans reflect the error on the fitted solar modulation parameters, cf.\ Tbl.~\ref{tab:solar}. Note that while the modulation to momenta $p_{\text{TOA}} > 1 \, \text{GeV}/c$ is constrained by AMS data, the modulation at lower momenta is an extrapolation. }
\label{fig:solmod_parameters_variation}
\end{figure*}

\begin{figure*}[t]
\centering
\includegraphics[width = 0.45\textwidth]{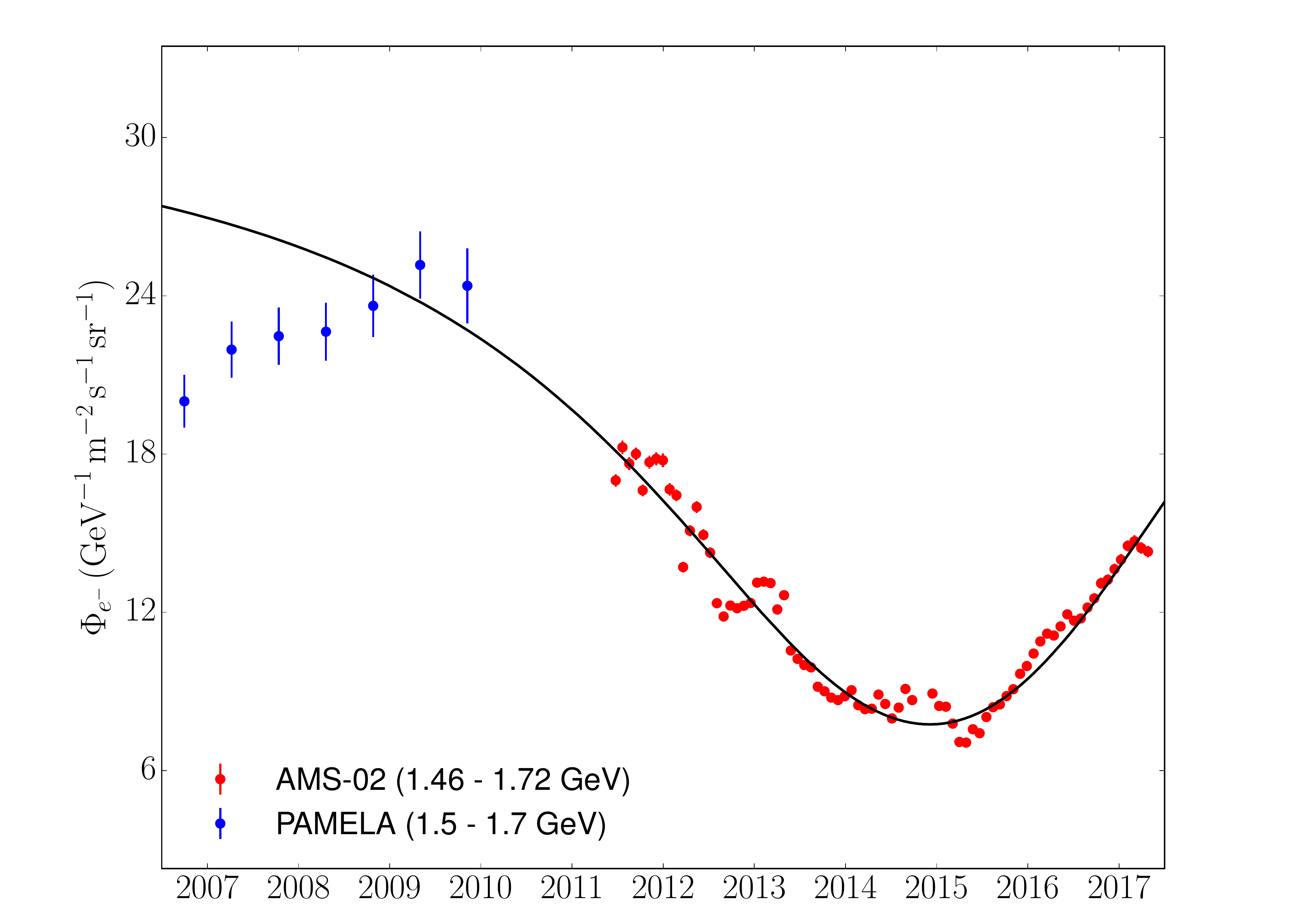}
\includegraphics[width = 0.45\textwidth]{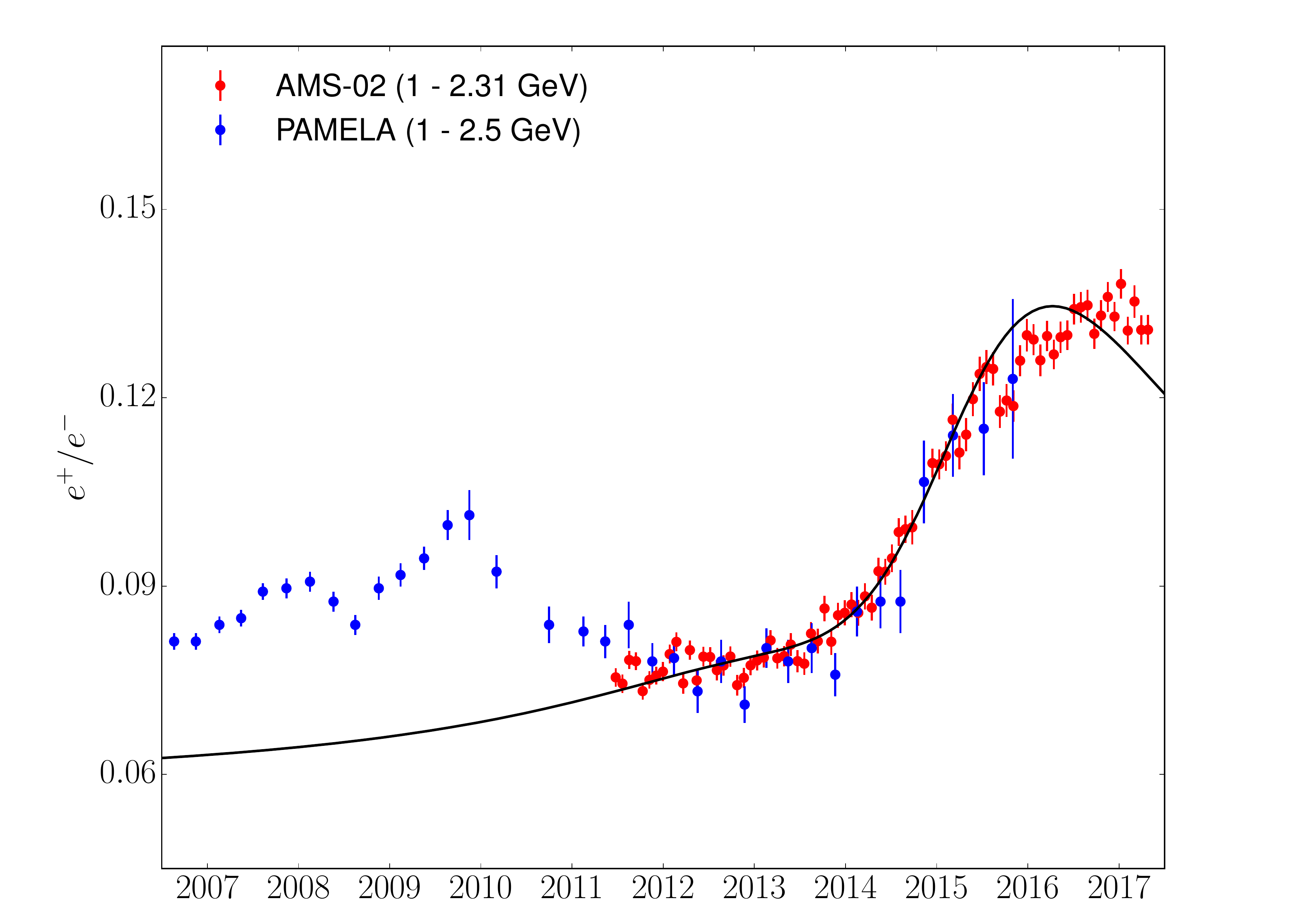}
\caption{Electron flux ({\it left panel}) and electron-to-positron
  ratio ({\it right panel}) in two different energy bins, as predicted
  from our model with the parameters of Table~\ref{tab:solar} compared
  with AMS data \cite{PhysRevLett.121.051102} (red points) and PAMELA data \cite{PhysRevLett.116.241105,Munini:2017ali} (blue points). In the
  case of the ratio, the AMS data points have been integrated over
  several energy bins, in order to have the maximal compatibility with
  the PAMELA energy bins.}
\label{fig:pamela}
\end{figure*}

\begin{figure*}[t]
\centering
\includegraphics[width = 0.49\textwidth]{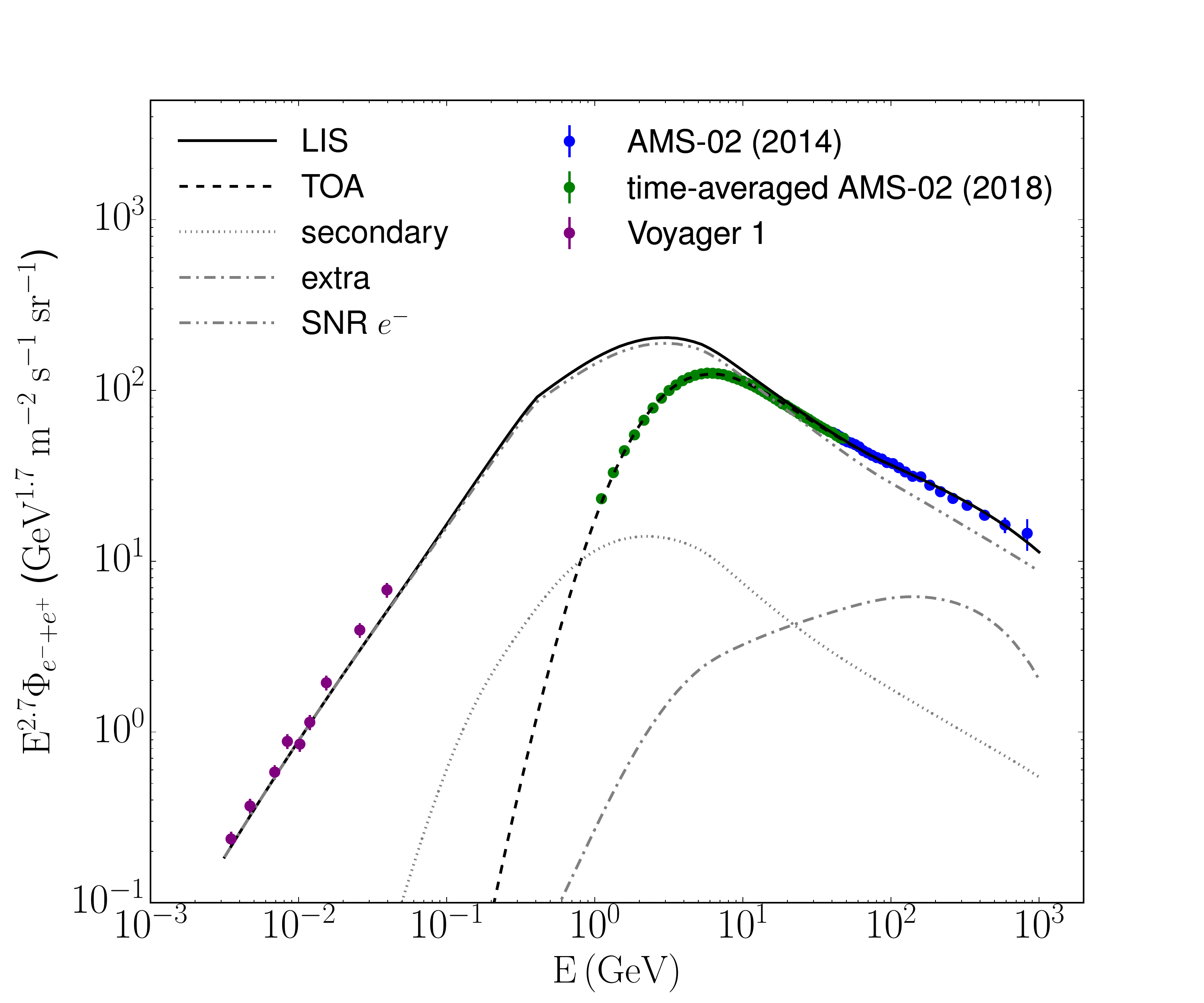}
\includegraphics[width = 0.49\textwidth]{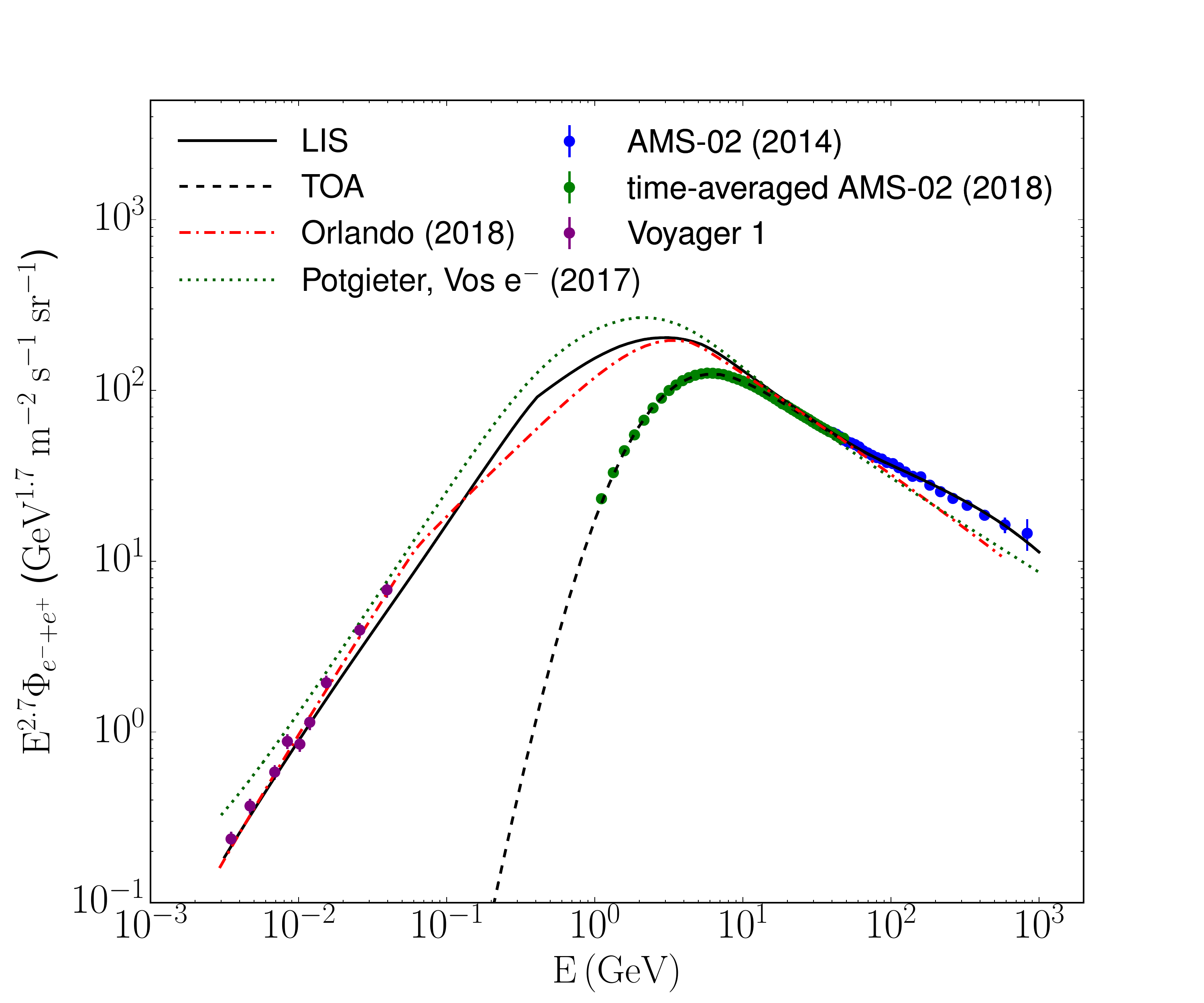}
\caption{{\bf Left}: All electron spectrum, plotted together with the different
  datasets considered in this work. The green points represent the
  average over the AMS data-taking period of the time-dependent
  datasets of \cite{PhysRevLett.121.051102}, while the dashed black
  line is the average over the same time period of the all-electron
  flux predicted by our solar modulation model with the best-fit
  parameters of Table~\ref{tab:solar}. The gray lines show the
  contributions from the different electron and positron sources. {\bf Right}: Same as in the left panel, with the addition of the $e^+ + e^-$ LIS of \cite{Orlando:2017mvd} (red, dot-dashed line) and the $e^-$ LIS from \cite{2017A&A...601A..23P} (green, dotted line).}
\label{fig:sum_alldata}
\end{figure*}

We use the local electron and positron fluxes measured by the AMS
experiment in the $1-50 \, \text{GeV}$ energy range. We model solar
modulation as described in Section~\ref{sec:solarmod_extended}.
We tune our model, which we label {\it 2 breaks model}, by performing a
global fit to all the data sets considered in the previous steps of
the analysis (radio, high-energy and Voyager data) together with the
time-dependent electron and positron data from AMS. In fitting the
high-energy electron and positron spectra we consider only data above
50 GeV, thus avoiding any overlap with the time dependent datasets. In
building the model, we start from the one described in the previous
section (i.e., with a low energy break in the electron spectrum in
order to reproduce Voyager data) and then we add the possibility to
have a second break in the electron spectrum at high energies (here
``high energies" means outside the regime where solar modulation has
an impact). This addition of a second break is motivated by our
expectation that a model able to correctly reproduce solar modulated
data will be characterised by an electron spectrum at
$\mathcal{O}$(1-10) GeV energies which will probably be softer than
the one found in the previous steps of the analysis. If the spectrum
is softer at low energies, a spectral hardening will be required to
reproduce the high energy end of the electron spectrum measured by
AMS. Adding a break adds two free parameters to the model, which now
consists of 9 parameters associated to the modelling of the positron
and electron LIS, in addition to the parameters associated to
the solar modulation model.

The results of the fit are reported in the bottom row of
Table~\ref{tab:bestfit_leptons}, while the various observables related
to the positron and electron LIS are shown in
Fig.~\ref{fig:results_fit} and the $\chi^2$ values associated to the
different data sets are listed in Table~\ref{tab:chi2_leptons}. In addition, the electron and positron LIS are reported in tabulated form in Appendix~\ref{sec:fluxes}. The
LIS are substantially modified now that solar modulated data are taken
into account. In particular, the electron LIS is significantly softer
at intermediate energies ($\Gamma_2$ changes from 2.57 to 2.69). This
requires a relatively strong hardening in order to fit high-energy
electron data. Overall, the fit to high-energy electron data is
significantly better than in the previous cases we
considered. Moreover,
as predicted, the fit to radio data worsens as a result of the
electron LIS being steeper at low energies, but it is still in an
acceptably good agreement with data. Another important consequence of the necessity to have a softer electron spectrum at $\mathcal{O}$(1-10) GeV energies is that the low-energy spectral break needed to fit Voyager data gets shifted to larger energies (it moves from 109 MeV to 411 MeV). This is illustrated in the bottom right panel of Fig.~\ref{fig:results_fit}. 

The results of the fit to the time-dependent AMS electron and positron
fluxes are shown in Figs.~\ref{fig:ele_solarmod} and
\ref{fig:pos_solarmod}, while the best fit solar modulation parameters
are reported in Table~\ref{tab:solar}. Our model provides a
satisfactory fit to the long-term behaviour of both the positron and
electron spectra, across the whole energy range covered by AMS
time-dependent data. However, it is also manifest that there are
several short-term variations in both spectra that cannot be described
within the framework of our model. Indeed, as discussed in
Section~\ref{sec:solarmod_extended}, our solar modulation model is
based on the assumption that the force-field potential varies smoothly
with time, which is not the case for the short-term events that appear
here. This complex structure of short-term fluctuations is the result
of different solar phenomena, namely coronal mass ejections and solar
wind streams, which can determine both increases (solar energetic
particle events) and decreases (Forbush decrease) of the fluxes of CRs
that reach Earth. These variations are typically much larger than the
experimental uncertainty associated to the AMS measurement of the
fluxes, in particular at the lowest energies. This is illustrated in
Figs.~\ref{fig:ele_solarmod} and \ref{fig:pos_solarmod} for the case
of the March 2012 Forbush decrease, which is one of the strongest
solar events recorded during the AMS data-taking period
\cite{doi:10.1002/2013JA019166}. The fact that the fluctuations in the spectra associated to
short term events are larger than the experimental uncertainties on
the data points is particularly true for the electron spectrum, which
is characterized by the smallest uncertainty and this explains the
large $\chi^2$ value associated to this dataset. Fig.~\ref{fig:ratio_solarmod}
illustrates the predicted $e^+/e^-$ ratio and it can be seen that
short-term events have a much more limited impact on the $e^+/e^-$
ratio. 

In order to better assess the performance of our solar modulation
model in fitting the average trend of the spectra, we compare in the left panel of 
Fig.~\ref{fig:solmod_parameters_variation} the maximum fluctuation of the electron
spectrum that our fit predicts with the one that is actually observed
in AMS data and with the fluctuation that results from a direct fit of
the spectrum with a harmonic function of period $T$:

\begin{equation}
\Phi_{e^-} (t) = A + B\,\mathrm{cos}\left( \frac{2\pi (t-t_0)}{T}\right). 
\label{eq:harmonic}
\end{equation} 

The variation predicted by our model is in good agreement with the one
resulting from the harmonic fit, which, albeit not being a physically
motivated model, certainly offers a good assessment of the maximum
fluctuations that can be found within the framework of a model where
the solar modulation parameters are assumed to have a smooth
dependence on time. The behavior of the harmonic fit at high energies
is that such fit is noise-dominated, because of the increased
uncertainty associated with AMS data. By comparing the fluctuations
observed in the data with the ones predicted by the harmonic fit and
by our model one can estimate the impact of short-term solar
events. The results shown in the left panel of Fig.~\ref{fig:solmod_parameters_variation} allow also for
an estimate of the maximum impact of solar modulation on the electron
flux: If one considers only the long-term fluctuations, such impact is still 
above 1\% at 20 GeV, while if one takes into account also short term
events, the impact reaches the 4\%.

In the right panel of Fig.~{\ref{fig:solmod_parameters_variation}} we show the unmodulated momenta $p_{\text{IS}}$ as a function of time for modulated momenta $p_{\text{TOA}} = 0.1$, $2$ and $10 \, \text{GeV}/c$, based respectively on eq.~(\ref{eq:phi_timedep}) and eq.~(\ref{eq:k_extFF}) with the parameters of Table~\ref{tab:solar}. Firstly, there is the usual overall trend with energy, in that high-energy electrons and positrons are less affected by solar modulation than low-energy electrons and positrons. (Compare for instance the $p_{\text{TOA}} = 0.1 \, \text{GeV}/c$ curves with the $p_{\text{TOA}} = 10 \, \text{GeV}/c$ ones.) Second, modulation is markedly charge-sign dependent: Electrons are modulated more strongly than positrons, in particular from $\sim$ 2014 onwards, whereas at earlier times both are modulated in similar ways. 
This is compatible with our
expectations as most of the AMS time-dependent data collected in
\cite{PhysRevLett.121.051102} refer to a period of positive polarity
of the heliospheric magnetic field. 
Indeed, as discussed in
\cite{Strauss2012}, because of drifts, 
positively charged particles at Earth have propagated across polar directions, while the negatively charged particles have travelled along the heliospheric current sheet (HCS). Due to the waviness of the HCS, negatively charged particles travel longer distances and are thus subject to stronger adiabatic losses. Our solar modulation model does not explicitly feature drift effects but we allow for different $\phi$ functions for electrons and positrons and thus the modulation can be different if the data require so. At earlier times, that is between $\sim$ 2011 and 2014, solar activity was at a maximum and electrons and positrons will have been modulated similarly. This is in line with the curves in the right panel of Fig.~{\ref{fig:solmod_parameters_variation}} being closer together.

Despite the fact that our solar modulation model is able to reproduce
remarkably well the average spectra, and with parameters that are
compatible with our expectations, a word of caution is in order about
the possibility to use the parameters of our model to make predictions
outside of the AMS data-taking period. Indeed, since our model is
based on a simplified description of solar modulation and, at present,
AMS data cover only a limited fraction of the solar cycle, we do not
expect our model to have a strong predictive power. To better
illustrate this point, we plot in Fig.~\ref{fig:pamela} the prediction
of our model, extended in the past by assuming a 22 years periodicity
for the electron and positron force-field potentials, compared to the
$e^-$ flux and the $e^+/e^-$ ratio measured by PAMELA
\cite{PhysRevLett.116.241105,Munini:2017ali}. The predictivity of our
solar modulation model will certainly improve if its parameters were
tuned on a dataset extended over a whole solar cycle, but still it has
to be taken into account that different solar cycles might also be
very different and therefore a simple periodicity of the force-field
potentials might not be a realistic assumption.

%------------------------------------------------------------------------------------------------------------------------------------------
%------------------------------------------------------------------------------------------------------------------------------------------

%------------------------------------------------------------------------------------------------------------------------------------------
%------------------------------------------------------------------------------------------------------------------------------------------

\section{Summary and outlook}
\label{sec:conclusions}

We have presented a model of the cosmic ray electron and positrons fluxes over a wide range of energies, from the MeV to the TeV domain, reproducing not only fluxes measured locally, but also the Galactic radio background and measurements outside the heliosphere.
As sources of electrons and positrons we have considered SNRs, charge-symmetric extra sources and spallation processes in the interstellar medium. Moreover, we have assumed the Galactic transport of electrons and positrons to be purely diffusive. 
In order to motivate the spectral breaks needed in the spectrum of SNRs, we have carefully considered the influence of different data sets. A satisfactory fit to the high-energy domain of the electron and positron fluxes measured by AMS and to the diffuse radio emission can be achieved with a simple power law for the electrons injected by SNRs. However, such a model overproduces electrons at $\mathcal{O}$(MeV) energies as measured by Voyager I.
Instead, a spectral break in the SNR spectrum is required.
Including in our fit also recent time-dependent electron and positron top-of-atmosphere fluxes, requires to model the effects of solar modulation, which we have performed within a simple extension of the standard force-field approximation. We have shown that the fit to solar modulated data requires the electron spectrum injected by SNRs to be steeper at $\mathcal{O}$(1-10) GeV energies and this makes a second spectral break at high energies necessary to fit the electron flux. In addition, the fit to the AMS time-dependent data sets has also proven that our solar modulation model works very well in reproducing the long-term trends of the low-energy electron and positron fluxes. 

A summary of the performance of our model in fitting the various data sets that we have considered in our analysis is shown in the left panel of Fig.~\ref{fig:sum_alldata}. In the plot we compare the time-averaged sum of the time-dependent electron and positron data sets with the prediction given by the {\it 2 breaks model}. The latter is obtained by averaging over the AMS data-taking period the flux that results from solar modulation, modelled within the extension of the force-field model presented in this paper, with parameters as in Table~\ref{tab:solar}. Data and theoretical prediction are in remarkable agreement. 

In the right panel of Fig.~\ref{fig:sum_alldata} we show the
comparison between the LIS predicted by our {\it 2 breaks model} and the LIS given in Refs.~\cite{Orlando:2017mvd} and \cite{2017A&A...601A..23P} (for this latter case we consider the $e^-$ LIS, as it is the only one provided in the paper). The three fluxes show a significant difference in the [100 MeV - 5 GeV] energy range. As it has been illustrated in this paper, this is the domain probed mostly by radio observations and by time-dependent solar modulated fluxes. It is worth mentioning that, among the three LIS considered here, our model is the only one that is tuned on both these observations, as the LIS from Ref.~\cite{Orlando:2017mvd} is not based on solar modulated data, while the model provided from Ref.~\cite{2017A&A...601A..23P} does not take into account radio constraints.   

% outlook
We hope that future studies will make use of these interstellar fluxes. Three applications seem most interesting and pressing: First, the inferred spectrum and charge symmetry of the extra component is tightly constrained by the measured electron flux even though it does not dominate the electron flux at any energy. Modifying the extra component would require modification of the electron spectrum at lower energies which again is constrained by a variety of data at these energies. Therefore, the extra component can be taken as a starting point for future studies of the origin of the positron excess. Second, at the highest energies, i.e. at TeV energies and beyond, stochasticity effects due to the discrete nature of the SNRs will shape the electron spectrum (e.g.~\cite{Mertsch:2018bqd}). As we were mainly concerned with lower energies, we have considered a smooth distribution of sources, which produces the expectation value of the flux at any one energy. This expectation value will be most valuable when investigating the possible origin of the break observed in the all-electron spectrum around a TeV~\cite{Aharonian:2009ah,Ambrosi:2017wek,KerszbergICRC2017}. Finally, what this study has also contributed is a new, effective way to take into account the effects of solar modulation in a time-dependent fashion. While our model has been fitted to time-dependent electron and positron fluxes above $\sim 1 \, \text{GeV}$, it should be easy to apply it to data sets extending to lower energies or the other species altogether.

\appendix
\section{Treatment of correlated uncertainties}
\label{sec:AMS_uncertainty}

The total uncertainty that characterizes an experimental measurement
at a given energy is given by the quadratic sum of the statistical
and systematic errors:
\begin{equation}
\sigma_{\mathrm{tot}} = \sqrt{\sigma_{\mathrm{syst}}^2 + \sigma_{\mathrm{stat}}^2}.
\end{equation}
When dealing with datasets that extend over an extended energy range,
the systematic uncertainty generally exhibits a certain degree of
correlation between different energy bins. This uncertainty has to be
taken into account when one is using the experimental uncertainties to
compute a $\chi^2$ in order to fit the data with a given theoretical
model.

The rigorous way of taking into account correlations requires the
knowledge of the correlation matrix, which unfortunately is not
available. Therefore, one has to resort to simpler recipes, such as
the one proposed by \cite{Cavasonza:2016qem}. Within such a framework,
one assumes that, when computing a $\chi^2$, the systematic
uncertainty of each data point is the sum of a fully correlated and a
fully uncorrelated component:
\begin{equation}
\sigma_{\mathrm{syst}} = \sigma_{\mathrm{syst,cor}}  + \sigma_{\mathrm{syst,unc}}, 
\end{equation}
with the uncorrelated component being 1\% of the measured value. Only
this uncorrelated component enters the definition of the total
uncertainty:
\begin{equation}
\sigma_{\mathrm{tot}} = \sqrt{ \sigma_{\mathrm{syst,unc}}^2 + \sigma_{\mathrm{stat}}^2}.
\end{equation} 
Concerning the correlated component, it can be treated as an overall
scale uncertainty on the acceptance which acts as an uncertainty on
the normalization of the measured quantity. This basically means that
the correlated component of the systematic uncertainty can be used to
determine an uncertainty on the values of the parameters of the
theoretical model that we are fitting against the data. Such
uncertainty is determined by fitting the data shifted upward and
downward by an amount that corresponds to the correlated uncertainty.

In the present work, we adopt the prescription described above in the
fit of the CR fluxes measured by AMS. On the contrary, when we fit
ratios between two CR species, we assume the correlated component of
the systematic uncertainty to be negligible,
i.e.~$\sigma_{\mathrm{syst}} = \sigma_{\mathrm{syst,unc}}$.

\section{Electron and positron LIS}
\label{sec:fluxes}
\setlength{\LTcapwidth}{0.5\textwidth}

\begin{longtable}{c c c}

\caption{Electron and positron LIS obtained within the {\it 2 breaks model}, as described in Section~\ref{sec:2breaks}. We limit the energy range to the [1 MeV - 1 TeV] interval, which is the one covered by the experimental data used in our analysis.}\\[6pt] \hline \\[3pt]
E  & $\Phi^{\mathrm{LIS}}_{e^-}  $  & $\Phi^{\mathrm{LIS}}_{e^+}  $  \\[2pt]
(GeV) &  (GeV m$^{2}$ s sr)$^{-1}$ &   (GeV m$^{2}$ s sr)$^{-1}$ \\[3pt] \hline \\ [2pt]
\endfirsthead

\multicolumn{3}{c}%
{{ \tablename\ \thetable{} -- continued from previous page}} \\[6pt] \hline\\[3pt]E  & $\Phi^{\mathrm{LIS}}_{e^-}  $  & $\Phi^{\mathrm{LIS}}_{e^+}  $  \\[2pt]
(GeV) &   (GeV m$^{2}$ s sr)$^{-1}$ &   (GeV m$^{2}$ s sr)$^{-1}$ \\[3pt] \hline \\ [2pt]
\endhead

\hline\\[2pt] \multicolumn{3}{c}{Continued on next page} \\[2pt] \hline
\endfoot

\hline
\endlastfoot
1.0000E-03  &	     4.0706E+06      &	         8.0054E+01 \\[3pt]
1.2589E-03  &	     3.1598E+06      &	         6.9219E+01 \\[3pt]
1.5849E-03  &	     2.4261E+06      &	         6.0430E+01 \\[3pt]
1.9953E-03  &	     1.8448E+06      &	         5.3926E+01 \\[3pt]
2.5119E-03  &	     1.3906E+06      &	         5.0008E+01 \\[3pt]
3.1623E-03  &	     1.0393E+06      &	         4.8814E+01 \\[3pt]
3.9811E-03  &	     7.7159E+05      &	         5.0640E+01 \\[3pt]
5.0119E-03  &	     5.6961E+05      &	         5.6308E+01 \\[3pt]
6.3096E-03  &	     4.1841E+05      &	         6.6900E+01 \\[3pt]
7.9433E-03  &	     3.0593E+05      &	         8.3202E+01 \\[3pt]
1.0000E-02  &	     2.2283E+05      &	         1.0529E+02 \\[3pt]
1.2589E-02  &	     1.6169E+05      &	         1.3209E+02 \\[3pt]
1.5849E-02  &	     1.1684E+05      &	         1.6068E+02 \\[3pt]
1.9953E-02  &	     8.4072E+04      &	         1.8885E+02 \\[3pt]
2.5119E-02  &	     6.0282E+04      &	         2.1606E+02 \\[3pt]
3.1623E-02  &	     4.3102E+04      &	         2.4114E+02 \\[3pt]
3.9811E-02  &	     3.0749E+04      &	         2.6021E+02 \\[3pt]
5.0119E-02  &	     2.1907E+04      &	         2.7088E+02 \\[3pt]
6.3096E-02  &	     1.5620E+04      &	         2.7206E+02 \\[3pt]
7.9433E-02  &	     1.1162E+04      &	         2.6299E+02 \\[3pt]
1.0000E-01  &	     7.9932E+03      &	         2.4085E+02 \\[3pt]
1.2589E-01  &	     5.7282E+03      &	         2.0806E+02 \\[3pt]
1.5849E-01  &	     4.1046E+03      &	         1.6834E+02 \\[3pt]
1.9953E-01  &	     2.9383E+03      &	         1.2784E+02 \\[3pt]
2.5119E-01  &	     2.0975E+03      &	         9.2251E+01 \\[3pt]
3.1623E-01  &	     1.4856E+03      &	         6.5237E+01 \\[3pt]
3.9811E-01  &	     1.0249E+03      &	         4.5712E+01 \\[3pt]
5.0119E-01  &	     6.4251E+02      &	         3.1521E+01 \\[3pt]
6.3096E-01  &	     3.9544E+02      &	         2.1044E+01 \\[3pt]
7.9433E-01  &	     2.4154E+02      &	         1.3525E+01 \\[3pt]
1.0000E+00  &	     1.4595E+02      &	         8.3521E+00 \\[3pt]
1.2589E+00  &	     8.6761E+01      &	         4.9468E+00 \\[3pt]
1.5849E+00  &	     5.0540E+01      &	         2.8252E+00 \\[3pt]
1.9953E+00  &	     2.8794E+01      &	         1.5665E+00 \\[3pt]
2.5119E+00  &	     1.6011E+01      &	         8.4622E-01 \\[3pt]
3.1623E+00  &	     8.6511E+00      &	         4.4531E-01 \\[3pt]
3.9811E+00  &	     4.5474E+00      &	         2.2935E-01 \\[3pt]
5.0119E+00  &	     2.3168E+00      &	         1.1543E-01 \\[3pt]
6.3096E+00  &	     1.1334E+00      &	         5.6334E-02 \\[3pt]
7.9433E+00  &	     5.3308E-01      &	         2.6776E-02 \\[3pt]
1.0000E+01  &	     2.4800E-01      &	         1.2782E-02 \\[3pt]
1.2589E+01  &	     1.1509E-01      &	         6.1871E-03 \\[3pt]
1.5849E+01  &	     5.3386E-02      &	         3.0430E-03 \\[3pt]
1.9953E+01  &	     2.4772E-02      &	         1.5210E-03 \\[3pt]
2.5119E+01  &	     1.1511E-02      &	         7.7192E-04 \\[3pt]
3.1623E+01  &	     5.3631E-03      &	         3.9751E-04 \\[3pt]
3.9811E+01  &	     2.5090E-03      &	         2.0749E-04 \\[3pt]
5.0119E+01  &	     1.1805E-03      &	         1.0950E-04 \\[3pt]
6.3096E+01  &	     5.5903E-04      &	         5.8203E-05 \\[3pt]
7.9433E+01  &	     2.6692E-04      &	         3.1015E-05 \\[3pt]
1.0000E+02  &	     1.2919E-04      &	         1.6476E-05 \\[3pt]
1.2589E+02  &	     6.2643E-05      &	         8.6909E-06 \\[3pt]
1.5849E+02  &	     3.0338E-05      &	         4.5425E-06 \\[3pt]
1.9953E+02  &	     1.4669E-05      &	         2.3462E-06 \\[3pt]
2.5119E+02  &	     7.0715E-06      &	         1.1910E-06 \\[3pt]
3.1623E+02  &	     3.3969E-06      &	         5.9272E-07 \\[3pt]
3.9811E+02  &	     1.6251E-06      &	         2.8800E-07 \\[3pt]
5.0119E+02  &	     7.7325E-07      &	         1.3580E-07 \\[3pt]
6.3096E+02  &	     3.6478E-07      &	         6.1251E-08 \\[3pt]
7.9433E+02  &	     1.7057E-07      &	         2.6371E-08 \\[3pt]
1.0000E+03  &	     7.9074E-08      &	         1.0779E-08 \\[3pt]
\end{longtable}
\clearpage

\bibliography{lowlep}

\end{document}